\documentclass[a4paper ,10pt]{article} 
\usepackage{graphicx,epsf}
\usepackage{amsbsy}

\begin{document}

\def\beq{\begin{equation}}
\def\eeq{\end{equation}}
\newcommand{\bea}{\begin{eqnarray}}
\newcommand{\eea}{\end{eqnarray}}
\def\bi{\begin{itemize}}
\def\ei{\end{itemize}}
\def\Tdot#1{{{#1}^{\hbox{.}}}}
\def\Tprime#1{{{#1}^{\hbox{'}}}}
\def\Tddot#1{{{#1}^{\hbox{..}}}}
\def\D{{\cal D}}
\def\d{{\delta}}
\def\T{{\bf T}}
\def\u{{\partial_u}}
\def\uc{u}
\def\qc{q}
\def\e{{=}}
\def\Dc{D}
\def\varphidc{\dot{ \tilde \varphi}{}}
\def\phid{\dot \phi}
\def\chid{\dot \chi}
\def\varphid{\dot \varphi}
\def\phidc{\dot{\tilde \phi}{}}
\def\chidc{\dot{\tilde \chi}{}}
\def\L{{\cal L}}
\def\R{{\cal R}}
\def\sigmadc{\dot{\tilde \sigma}{}}
\def\sigmad{\dot \sigma}
\def\Pdc{\dot{\tilde P}{}}
\def\rhodc{\dot{\tilde \rho}{}}
\def\betadc{\dot{\tilde \beta}{}}
\def\alphadc{\dot{\tilde \alpha}{}}
\def\HH{{\cal H}}
\def\ut{\tilde u}
\def\qt{\tilde q}
\def\ht{\tilde h{}}
\def\Dt{\tilde D}
\def\hc{h}
\def\I{I}
\def\It{{(I)}}
\def\J{J}
\def\Jt{{(J)}}
\def\p{\varphi}
\def\vp{{\boldsymbol{\varphi}}}
\def\e{{\boldsymbol{e}}}

\def\besi{\bar{e}_{\sigma}}
\def\bes{\bar{e}_{s}}

\def\Xf{X^{(1)}}
\def\Xs{X^{(2)}}
\def\af{{\delta \alpha}}
\def\as{{\delta \alpha}^{(2)}}
\def\sif{{\delta \sigma}}
\def\sis{{\delta \sigma}^{(2)}}
\def\sf{{\delta s}}
\def\ss{{\delta s}^{(2)}}
\def\phf{{\delta \varphi}}
\def\phs{{\delta \varphi}^{(2)}}
\def\phif{{\delta \phi}}
\def\chif{{\delta \chi}}
\def\sif{{\delta \sigma}}
\def\phis{{\delta \phi}^{(2)}}
\def\rhof{{\delta \rho}}
\def\rhos{{\delta \rho}^{(2)}}
\def\ps{{\delta P}^{(2)}}
\def\pf{{\delta P}}
\def\ab{\bar{\alpha}}
\def\rhob{\bar{\rho}}
\def\pb{\bar{P}}
\def\sib{\bar{\sigma}}
\def\phib{\bar{\phi}}
\def\chib{\bar{\chi}}
\def\sb{\bar{s}}
\def\thetab{\bar{\theta}}
\def\tackr{&\!\!\!}
\def\tackl{&\!\!\!}

\def\Zs{Z_{(s)}}

\def\Zsi{Z_{(\sigma)}}


\def\endignore{}
\def\ignore #1\endignore{} 
\def\Box{{\hbox{$\sqcup$}\llap{\hbox{$\sqcap$}}}}
\def \lsim{\mathrel{\vcenter
     {\hbox{$<$}\nointerlineskip\hbox{$\sim$}}}}
\def \gsim{\mathrel{\vcenter
     {\hbox{$>$}\nointerlineskip\hbox{$\sim$}}}}

\newcommand{\sfrac}[2]{{\textstyle\frac{#1}{#2}}}
\def\be{\begin{equation}}
\def\ee{\end{equation}}
\def\bea{\begin{eqnarray}}
\def\eea{\end{eqnarray}}
\def\nn{\nonumber}
\def\ansatz{{\it ansatz}}
\def\ansatze{{\it ans\"atze}}
\def\exd{{\rm d}}
\def\pref#1{(\ref{#1})}
\def\endignore{}
\def\ignore #1\endignore{} 

\def\bd{\begin{displaymath}}
\def\ed{\end{diplaymath}}

\def\eg{{\it e.g.}}
\def\ie{{\it i.e.}}
\def\vev{{\it vev}}
\def\d{\mathrm{d}}

\def\cW{{\cal W}}
\def\cA{{\cal A}}
\def\cR{{\cal R}}
\def\cN{{\cal N}}
\def\cF{{\cal F}}
\def\cK{{\cal K}}
\def\cT{{\cal T}}
\def\cL{{\cal L}}

\def\O{\mathcal{O}}
\def\H{\mathcal{H}}
\def\hO{\hat{O}}
\def\C{\mathcal{C}}
\def\L{\mathcal{L}}
\def\K{\mathcal{K}}
\def\F{\mathcal{F}}
\def\dd{\left|\partial d\right|}
\def\del{\nabla}
\def\Et{\tilde{E}}
\def\Bt{\tilde{B}}
\def\nn{\nonumber}
\def\cosech{\mathrm{cosech}}
\def\cosec{\mathrm{cosec}}
\def\x{\mathbf{x}}
\def\d{\mathrm{d}}
\def\yt{\tilde{y}}
\def\mn{_{\mu \nu}}
\def\mupn{^\mu_{\, \nu}}
\def\({\left(}
\def\){\right)}
\def\eps{\epsilon}
\def\hg{\hat{g}}
\def\k{\kappa}

\def\hW{{\hat{W}}}
\def\hN{{\hat{N}}}
\def\hR{{\hat{R}}}
\def\hA{{\hat{A}}}

\def\uv{\mu \nu}
\def\nab{\bar{\eta}_a}
\def\nbb{\bar{\eta}_b}
\def\pab{\bar{\varphi}_a}
\def\na{\hat{\eta}_a}
\def\la{\lambda_a}
\def\ta{\tau_a}

\def\n{\eta}
\def\l{\lambda}

\def\hcW{{\hat{\cal W}}}
\def\hcN{{\hat{\cal N}}}
\def\hcR{{\hat{\cal R}}}
\def\hcA{{\hat{\cal A}}}

\def\bea{\begin{eqnarray}}
\def\eea{\end{eqnarray}}
\def\be{\begin{equation}}
\def\ee{\end{equation}}

\def\exd{{\rm d}}
\def\cosh{{\rm cosh}}
\def\sinh{{\rm sinh}}
\def\tanh{{\rm tanh}}
\def\e{{\rm e}}

\def\d{ {\rm d} }
\def\hf{\frac{1}{2}}

\def\rpm{r_\pm}
\def\rp{r_+}
\def\rmm{r_-}
\def\rt{\tilde{r}_\pm}
\def\bd{\boxdot}
\def\rrp{\tilde{r}_+}
\def\rrm{\tilde{r}_-}

\def\PX{P_{,X}}
\def\p{\phi}
\def\s{\sigma}
\def\Z{Z_{(\s)}}
\def\W{Z_{(s)}}
\def\bc{\bar{c}_s}
\def\M{M}
\def\bP{\bar{P}}

\def\X{\xi}

\def\pr{p}
\def\Du{{\cal D}_u}
\def\Da{{\cal D}_{\perp a}}

\newcommand{\ind}[2]{\!\!\!\phantom{#1}^{(#2)}\!#1}

\begin{center}
{\LARGE \bf
Nonlinear perturbations of cosmological
scalar fields\\
  with non-standard kinetic terms}\\[1cm]

{
{\large\bf 
 S\'ebastien Renaux-Petel$^{a\,}$\footnote{E-mail: 
  renaux@apc.univ-paris7.fr}\,\,
  and\,\, Gianmassimo Tasinato$^{b\,,c\,}
 \footnote{E-mail: 
  g.tasinato@thphys.uni-heidelberg.de}$
}
}
\\[7mm]
{\it $^a$ 
APC (Astroparticules et Cosmologie),\\
UMR 7164 (CNRS, Universit\'e Paris 7, CEA, Observatoire de
Paris)\\
 10, rue Alice Domon et L\'eonie Duquet,
 75205 Paris Cedex 13, France
}\\[3mm]
{\it $^b$ Institut f\"ur Teoretische Physik, Universit\"at Heidelberg, 
Philosophenweg 16 and 19,\\
D-69120 Heidelberg, Germany
}\\[3mm]
{\it $^c$ Instituto de Fisica Teorica, UAM/CSIS Facultad de Ciencias C-XVI, \\
C.U. Cantoblanco, E-28049-Madrid, Spain
}
\\[1cm]
\vspace{-0.3cm}

\vspace{1cm}

{\large\bf Abstract}

\end{center}
\begin{quote}

{
We adopt a covariant
formalism to derive exact evolution equations for nonlinear perturbations, in
a universe dominated by two scalar fields. These scalar fields
are characterized by non-canonical kinetic terms and an
arbitrary field space metric, a situation typically encountered
in inflationary models inspired by string theory.
 We decompose the nonlinear scalar perturbations  into
 adiabatic and entropy modes, generalizing the definition adopted
 in the linear theory, and we derive the corresponding exact
 evolution equations.  We also obtain a nonlinear generalization
 of the curvature perturbation on uniform density hypersurfaces,
 showing that on large scales it is sourced 
 only by the nonlinear version of the entropy perturbation.  
 We then expand these equations to second
 order in the perturbations, using a coordinate based formalism. Our 
 results are relatively compact and elegant and enable one to identify
 the new effects coming from the non-canonical structure of the scalar fields Lagrangian.
 We also explain  how to analyze, in our formalism,  the interesting scenario of multifield Dirac-Born-Infeld inflation.
  }

\end{quote}

\newpage

\section{Introduction}

Inflation offers a compelling mechanism to produce a spatially flat and
approximately homogeneous universe. Moreover, it also provides a source
for small primordial fluctuations that seed the observed large scale
structures.  Simple models of inflation are realized in terms of a single,
slowly rolling scalar field characterized by canonical kinetic terms and
a flat potential. In this case, the spectrum of fluctuations  
 is characterized by  adiabatic, almost scale invariant
 density perturbations,  with nearly
 Gaussian distribution on
super-horizon scales.

While present-day observations are consistent
with  these models, there are good
theoretical reasons 
to go beyond the simple hypotheses on which they are based. As an
 important example, string theory motivates 
frameworks in which many scalar fields, with non-canonical kinetic terms and not necessarily flat potential, play a role
during inflation.  Indeed, string theory  
predicts the existence of a large number of scalar fields, the moduli, which in the early universe may be sufficiently light to interact with the
inflaton field, playing an active role during inflation.  Moreover,
   the
most studied stringy inflationary models are based on the dynamics of
D-branes moving in higher dimensional spaces:
in these set-ups,   the inflaton field is    
governed by a Dirac-Born-Infeld (DBI)  action characterized
by  non-canonical kinetic
terms (see \cite{McAllister:2007bg} for
recent reviews). In some regimes, the non-canonical
kinetic terms allow one to obtain inflation also with 
steep scalar potentials,
 as in the DBI-inflationary models of \cite{Silverstein:2003hf}-\cite{Chen:2005ad}. A more general
 analysis  of 
the spectrum of primordial fluctuations  for  multi-field
models with non-canonical
 kinetic terms, then, may allow one to understand how
future observations will be able to probe, or exclude,
inflationary models motivated or
inspired   by string theory.

\smallskip

Single-field 
inflationary models with non-canonical kinetic terms, of a class belonging to models of  \textit{k}-inflation \cite{ArmendarizPicon:1999rj},  have 
received much attention in the past 
years, for the possibility to  generate spectra
of fluctuations  with 
 non-Gaussian features,  
at a level that can be probed by future observations (see
 \cite{Chen:2006nt} for
 review and a comprehensive analysis of this case). More
recently, models of inflation with more than one scalar
field with non-canonical kinetic terms have been also considered
\cite{Easson:2007fz}-\cite{Cai:2008if}.  In this case, non-standard
correlations between adiabatic and entropy modes can
 occur,
and this affects the amplitude of non-Gaussianities 
in these models \cite{Langlois:2008qf}-\cite{Arroja:2008yy}. These correlations can even persist in the primordial radiation dominated era, as first pointed out in \cite{Langlois:1999dw}. The analysis of the examples 
 studied so far suggests that the dynamics of fluctuations,
 in string  inspired multi-field inflationary models, have
 new and rich features, depending on the  form
 of kinetic terms for the inflaton field(s).

\smallskip

In this paper, motivated
by the previous arguments,
 we analyze cosmological fluctuations of 
a system of two
scalar fields with non-canonical kinetic terms  at a 
fully  nonlinear level. We  adopt the covariant formalism
developed in \cite{Langlois:2005ii}-\cite{Langlois:2006vv}, along the
lines of earlier works by Ellis and Bruni \cite{Ellis:1989jt} and Hawking
\cite{Hawking:1966qi} (see also \cite{Bruni:1991kb}).
This formalism is particularly suitable to analyze
  fluctuations of the system we consider, since 
  it leads to
  a clear and natural decomposition of fluctuations in adiabatic
   and entropy components. This separation has
   been first studied, at the linear level, in 
  \cite{Gordon:2000hv} 
   for the case of canonical kinetic terms
  (see \cite{Wands:2007bd} for a recent review). It 
is particularly convenient for analyzing the generation
and conversion of adiabatic and isocurvature components
of fluctuations produced 
during inflation. It also plays an important
role in the previously
mentioned recent papers, that analyze   fluctuations for multi-field
models with non-canonical kinetic terms.

Following \cite{Langlois:2006vv}, 
we  generalize this decomposition into adiabatic and isocurvature components to a fully nonlinear
set-up. In order to do so, we first define adiabatic and
isocurvature covectors, representing the nonlinear generalization
of linear fluctuations, in the case of general field space metric.
Then, we 
 obtain a set of {\it exact} evolution equations
for these quantities, in our framework characterised by a non-standard Lagrangian. In addition,
 by defining a suitable nonlinear
 generalization of the curvature perturbation, we  derive 
  the evolution equation for this quantity and we 
 determine 
 how
  it  is sourced by the nonlinear entropy covector.
   This result 
  generalizes  already known linear equations
   that
  play an important role for analyzing
    how entropy fluctuations  
  are converted into adiabatic  modes on superhorizon scales.

We then show how our results
   can  be reexpressed in a more familiar coordinate based 
   approach,  and  how our equations can be
   expanded 
   to first and  second order. Proceeding in this way, we
   re-obtain at first order
     the linear evolution equations determined in 
     \cite{Langlois:2008mn}, while
   we  find new results when pursuing the expansion up
    to second order.   In particular, working in a large scale limit,
we determine how the curvature perturbation is sourced
by the entropy modes at second order. Our final equations
are relatively 
compact, and allow one to clearly appreciate the
impact of non-canonical kinetic terms on the evolution 
of fluctuations. 

\smallskip
Our discussion closely follows \cite{Langlois:2006vv},
to which we refer the reader for further details and 
references. We show how to extend the methods
  of this paper 
 to the case of non-canonical kinetic terms
 for the scalar fields, 
 and for general field space metric.  
 In doing so, we obtain 
  nonlinear equations  for  the fluctuations, 
  that can be expressed
   in a  physically transparent way,
and we 
 show  the usefulness of the covariant approach for analyzing  cosmological models inspired by string theory.

\section{Covariant formalism for  scalars
with non-canonical kinetic terms}{\label{sec:general}}

We consider an arbitrary unit timelike vector $u^a= d
x^a/d \tau$ ($u_a u^a =-1$),  defining a congruence of
cosmological observers. The spatial projection tensor orthogonal
to the four-velocity $u^a$ is provided by
\beq
h_{ab}\equiv g_{ab}+u_a u_b, \quad \quad (h^{a}_{\ b} h^b_{\
c}=h^a_{\ c}, \quad h_a^{\ b}u_b=0).
\eeq
To describe the time evolution, the covariant definition of
the time derivative    will be the {\em Lie derivative}
with respect to $u^a$, defined for a generic covector
$Y_a$ by (see e.g. \cite{wald}) 
\begin{equation}
\dot Y_a\equiv\L_u Y_a \equiv u^b \nabla_b Y_a + Y_b \nabla_a u^b,
\label{Lie}
\end{equation}
and will be denoted by {\em a dot}. 
 For scalar  quantities, one simply has
\beq
\dot f = u^b \nabla_b f.
\eeq

To describe perturbations in the covariant approach, we consider
the projected covariant derivative orthogonal to the four-velocity
$\uc^a$, denoted by $\Dc_a$. For a
generic tensor, the definition is
\beq
D_a T_{b\dots}^{\ c\dots}\equiv h_a^{\ d}h_b^{\ e}\dots h^{\
c}_f\dots\nabla_d T_{e\dots}^{\ f\dots}.
\eeq
In particular, when focussing on a scalar quantity $f$,
this reduces to
\beq
\Dc_a f\equiv h_a^{\ b} \nabla_b f = \partial_a f + u_a \dot f \, .
\label{def_Daf}
\eeq
We can also   decompose
\beq
\nabla_b u_a=\sigma_{ab}+\omega_{ab}+{1\over 3}\Theta h_{ab}-a_a
u_b, \label{decomposition}
\eeq
with the (symmetric) shear tensor $\sigma_{ab}$ and the
(antisymmetric) vorticity  tensor $\omega_{ab}$; the volume
expansion, $\Theta$, is defined by
\beq
\Theta \equiv \nabla_a u^a,
\eeq
while
\beq
a^a \equiv u^b\nabla_b u^a
\eeq
is the acceleration vector.

\smallskip

Let us now  consider $\rm N$ scalar fields
minimally coupled to gravity with general Lagrangian density (see \cite{Langlois:2008mn} for a study of the linear perturbations in this type of model):
\beq
{\cal L} = P(X, \phi^I)\,.
\label{lagrangian}
\eeq
Here $X$
is 
\bea
X&=& -\frac12 \, G_{IJ} \, g^{a b} \, \nabla_a \phi^I 
\nabla_b \phi^J\,,
\eea
or, using Eq. (\ref{def_Daf}), 
\bea
X&=&\frac12\,G_{IJ}  \dot{\p}^I  \dot{\p}^J  -\frac12 G_{IJ} D_a \phi^I
D^a \phi^J
\label{Xdecomposition}
\eea
and $G_{IJ}\equiv G_{IJ}(\phi^K)$ is a metric in `field space' that
can be used to  raise and lower field indices, denoted
by capital letters.  One can consider more general Lagrangians,
like the ones  studied in $\cite{Langlois:2008qf,Arroja:2008yy}$. 

However, the choice of Lagrangian density Eq.~(\ref{lagrangian}) is suitable
 for describing most of 
 \textit{k}-inflationary models
 considered in the literature. Especially, it is important
   to point out that  our formalism can be  applied 
   to two-field DBI inflationary  models of the type studied in \cite{Langlois:2008wt,Langlois:2008qf}. One of the present author (S. RP) has shown in these works that the multi-field DBI Lagrangian can \textit{not} be written in the form $P(X, \phi^I)$. It can however be written in the form  $P(\tilde{X}, \phi^I)$ where $X$ and $\tilde{X}$ only differs by terms in spatial gradients. Therefore, all the results derived in this paper \textit{in the large scale limit}, where spatial gradients can be neglected, are readily applicable to multi-field DBI models simply by considering the specific Lagrangian 
 \beq
 P(X,\p^I)=-\frac{1}{f(\p^I)} \left(\sqrt{1-2 f(\p^I) X}-1 \right)-V(\p^I)
 \label{DBI}
 \eeq
 where $f$ is the so-called warp factor and $V$ is a potential term (see \cite{Langlois:2008qf} for more details on multi-field DBI
 models).

\smallskip

 The
 energy momentum tensor derived from Eq.~(\ref{lagrangian}) reads 
 \be
T_{ab} = P_{,X} G_{IJ} \nabla_a \phi^I \nabla_b \phi^J
+ g_{ab}\,P\,.
 \label{EMT}
\ee

Given an arbitrary unit timelike vector field $u^a$, it is always
possible to decompose the total energy momentum tensor as
\beq
T_{ab} =(\rho+\pr) u_a u_b  + q_au_b+ u_aq_b+ g_{ab}\,\pr
 + \pi_{ab}, \label{EMT2}
\eeq
where $\rho$, $\pr$,
 $q_a$ and $\pi_{ab}$ are respectively the energy density,
pressure,  momentum  and anisotropic stress tensor measured in the
frame defined by $u^a$.
Starting from the expression for
the  energy-momentum tensor (\ref{EMT}) one finds 
(we write $\dot{\p}_I\,\equiv\,G_{IJ}\,\dot{\p}^J$)
\bea
\rho \tackl \equiv \tackr 
T_{ab}u^au^b= 
P_{,X}\,  \dot{\p}^I  \dot{\p}_I -P\,\,
, \label{rhotot}\\
\pr
 \tackl \equiv \tackr
  \frac{1}{3}h^{ac}T_{ab}h^b_{\, c}= 
\frac13 P_{,X} G_{IJ} D_a \phi^I 
D^a \phi^J\,+\,P 
, \label{Ptot}\\
q_a \tackl \equiv \tackr 
-u^bT_{bc}h^c_{\, a}\,=\,  
 - P_{,X} \dot{\phi}_I
D_a \phi^I
, \label{q}\\
\pi_{ab} \tackl \equiv \tackr   h_a^{\, c}T_{cd} h^d_{\, b}- p\, h_{ab}= 
P_{,X}\,\left(
G_{IJ} D_a \phi^I D_b \phi^J-\frac{h_{ab}}{3}\,
G_{IJ}D_c \phi^I D^c \phi^J 
\right)
.
\label{as}
\eea

 \smallskip
 
 The evolution equations for the scalar fields are obtained from the variation
of the action with respect to the fields themselves. 
In our case one gets

\be\label{geneq}
P_{,X}\, G_{IJ} \nabla^a 
\nabla_a \phi^J+P_{,X}\,\Gamma_{IJK} \left( \nabla_a \phi^J \right) \left(\nabla^a \phi^K \right)+ \left(\nabla^{a}
P_{,X } \right) \,G_{IJ} \nabla_a \phi^J
+P_{,I}\,=\,0
\ee
where $\Gamma_{IJK} \equiv G_{IL} \Gamma^L_{JK} \equiv \frac 1 2 \left(G_{IJ,K}+G_{IK,J}-G_{JK,I} \right)$ is the Christoffel symbol associated to the metric $G_{IJ}$. The previous equation reduces to the
usual one, in the case one chooses
$P=X-V(\p^I)$, where   $V$ is a potential.

\smallskip

Notice that using the equalities, following from (\ref{def_Daf}) and (\ref{decomposition}),
\be
D_a D^a \,\phi^I
\,=\, \nabla_a \nabla^a \phi^I
 +\ddot{\phi}^I
 + \Theta \,\dot{\phi}^I
 -a^{b}\,D_b \phi^I\,,
\ee
\be
 \nabla^{a}
P_{,X } \,\nabla_a \phi^J\,=\, D^{a}
P_{,X } \,D_a \phi^J
-\dot{P}_{,X } \,\dot{\phi}^J\,,
\ee
\be
 \nabla^{a}
\p^J \,\nabla_a \phi^K\,=\, D^{a}
\p^J \,D_a \phi^K
-\dot{\p}^J \,\dot{\phi}^K\,,
\ee
equation (\ref{geneq}) becomes
\be\label{homgeneq}\label{evol_single}
\label{evolp}
 \ddot{\phi}^I\,+\, \Gamma^I_{JK} \,\left( \dot{\p}^J \dot{\p}^K-D_a \p^J D^a \p^K \right)\,+\,\left( \Theta+\frac{\dot{P}_{,X}}{{\PX}} \right)\, \dot{\p}^I
-\,\frac{1}{\PX}\,G^{IJ}P_{,J}
\, -\, D_a D^a \phi^I\,-\, a^b D_b \phi^I
\,-\, \frac{\left( D^{b} P_{,X}\right)}{\PX}\,D_b \phi^I
\,=\,0\,.
\ee
To simplify the notation, 
it is useful to define the spacetime derivative of field space vectors in curved coordinates
\beq
{\cal D}_a A^I \equiv \nabla_a A^I + \Gamma^I_{JK} \nabla_a \p^J A^K\,,
\eeq
with which we define  a time derivative in field space
\beq
\Du A^I\equiv u^a {\cal D}_a A^I
\eeq
and a spatially projected derivative in field space
\beq
\Da T_{b\dots}^{\I\, c\dots}\equiv h_a^{\ d}h_b^{\ e}\dots h^{\
c}_f\dots {\cal D}_d T_{e\dots}^{\I\, f\dots}\,.
\eeq
Notice that 
$\mathcal{D}_a$ acts as an ordinary time derivative on field space scalars (i.e. quantities without field space indices) and $\mathcal{D}_a G_{IJ}=0$.
 Using these definitions, it is possible to 
rewrite (\ref{evolp}) in a more condensed form as
\beq
\label{compact-evol-p}
\Du \dot{\p}^I \,+\,\left( \Theta+\frac{\dot{P}_{,X}}{{\PX}} \right)\, \dot{\p}^I
-\,\frac{1}{\PX}\,G^{IJ}P_{,J}
\, -\, \Da \left( D^a \phi^I \right) \,-\, a^b D_b \phi^I
\,-\, \frac{\left( D^{b} P_{,X}\right)}{\PX}\,D_b \phi^I
\,=\,0\,.
\eeq

\section{Single field}

We start our discussion examining the simple situation of
a single field. This provides  the opportunity to introduce some quantities
that will play an important role in the following discussion. 

Let us denote by $\p$ the  single  scalar field we are considering and set the trivial field space metric $G_{11}=1$.
 Then  the Klein-Gordon equation (\ref{compact-evol-p}) reads
\beq
\label{single-field}
\ddot{\p}\,+\,\left( \Theta+\frac{\dot{P}_{,X}}{{\PX}} \right)\, \dot{\p}
-\,\frac{1}{\PX}\,\frac{d P}{d \p}
\, -\, D_{a}  D^a \phi \,-\, a^b D_b \phi
\,-\, \frac{\left( D^{b} P_{,X}\right)}{\PX}\,D_b \phi
\,=\,0\,.
\eeq
By choosing 
\be
u^a_{com}\,=\,
\pm\frac{\nabla^a \phi}{\sqrt{-
\nabla_a \phi \nabla^a \phi
}}\label{u_comoving}
\ee
one finds $D_a \phi =0$. Then 
  (\ref{single-field}) becomes formally identical to the 
  well known homogeneous version,
    although it remains fully inhomogeneous and nonlinear. 

\smallskip

One can proceed  further, and  derive an exact and covariant
equation that mimics the equation of motion governing the linear
perturbations of a scalar field in a perturbed FLRW (Friedmann-Lema\^itre-Robertson-Walker) spacetime. The
idea  is to  consider  the evolution of the  space-time
 gradient  of the
scalar field, i.e.,  the  covector
\beq
\phi_a \equiv \nabla_a \phi.  \label{phi_a_def}
\eeq
Indeed 
this quantity can be decomposed into a spatial gradient and a
longitudinal component,
\beq
\phi_a \, =\, -\dot \phi\  u_a+ \Dc_a\phi.
\eeq
resembling, within   our full nonlinear
setting, the usual decomposition   of a quantity in a homogeneous
part, plus a perturbation.   
Notice  that when making  the particular
 choice (\ref{u_comoving}) the
spatial gradient disappears in the above expression,
and indeed the evolution equation becomes in this case
formally identical to the homogeneous one.

We then construct a second-order (in time) evolution equation for
$\phi_a$. We recall  that a dot stands for  the Lie derivative
along $u^a$, as defined in Eq.~(\ref{Lie}). We can derive the
evolution equation for $\phi_a$ by taking the
 spacetime gradient of Eq.~(\ref{single-field}) and noting that, for
any scalar $\phi$,  the Lie derivative with respect to $u^a$ and
the spacetime gradient (but not the spatial gradient) commute,
i.e.,
\beq
\nabla_a \dot \phi = \Tdot{(\nabla_a \phi)}.
 \label{prop_Lie}
\eeq
Acting with the space-time gradient $\nabla_a$
on equation (\ref{single-field}), and using the previous
definitions, we obtain

\begin{eqnarray}
\label{eq_phi_a} 
&&\ddot \phi_a +\left( \Theta\,+\frac{\dot{P}_{,X}}{\PX}\right)
 \dot \phi_a +\dot{\p} \nabla_a \left( \frac{\dot \PX}{\PX} \right)- \left[ \frac{1}{\PX}\,\frac{d P}{d \p} \right]_{,X} \nabla_a X - \left[ \frac{1}{\PX}\,\frac{d P}{d \p} \right]_{,\p} \p_a  \cr && \hskip 0.5cm
  =\,\,-\dot \p\, \nabla_a \Theta
  +  \nabla_a \left(
 D_b D^b \phi\right)
+\, \nabla_a 
\left(a^{b} D_{b} \phi \right)
+\nabla_a 
  \left( \frac{D^{b}{\PX}}{ P_{,X}} D_b \phi\right)
  \end{eqnarray}
This equation is similar
to  the analogous perturbation equation at linear
order,
 but it additionally   incorporates the fully nonlinear dynamics
  of the
scalar field perturbation.

\subsection{Integrated expansion perturbation on co\-mo\-ving sli\-ces}

One
can define a covariant generalization of  the comoving
curvature perturbation using appropriate combinations of
spatially projected gradients. For a scalar field, a natural
choice is the covariant integrated expansion perturbation on
comoving hypersurfaces
$\R_a$, defined as
\beq
\R_a \equiv -D_a \alpha + \frac{\dot \alpha}{\dot \phi} D_a \phi,
\label{R_def}
\eeq
where $\alpha$ is the integrated volume expansion along $u^a$,
\beq
\alpha \equiv {1\over 3}\int d\tau \, \Theta \quad \quad (\Theta
= 3 \dot \alpha  ) \label{alpha_def}.
\eeq
 Since $\Theta/3$ corresponds to the local Hubble parameter, one sees that the quantity $\alpha$ can
be interpreted as the number of e-folds measured along the world-line of a cosmological observer with
four-velocity $u^a$.

When the four-velocity $u^a$ is chosen to be comoving with the scalar field, as defined in Eq.~(\ref{u_comoving}), then
the last term in the definition (\ref{R_def}) drops out. 

Note a useful  property of  $\R_a$: one can
replace in its definition (\ref{R_def})
the spatial gradients $D_a$ by partial or covariant
derivatives,
\beq
\R_a = -\nabla_a \alpha + \frac{\dot \alpha}{\dot \phi} \phi_a.
\eeq
The same property applies to  the nonlinear
generalization of the comoving Sasaki-Mukhanov variable for a
scalar field, which can be defined as
\beq
Q_a \equiv D_a \phi - \frac{\dot \phi}{\dot \alpha} D_a \alpha
=\frac{\dot \phi}{\dot \alpha}\R_a. \label{Q_single}
\eeq

\subsection{Integrated expansion perturbation on uniform energy density slices}

Following \cite{Langlois:2005ii,Langlois:2005qp},
it is also possible to generalize the curvature perturbation on
uniform energy density hypersurfaces.
 The key
role is then played by the covector $\zeta_a$ defined as
\beq
\zeta_a \equiv D_a \alpha - \frac{\dot \alpha}{\dot \rho} D_a
\rho. \label{zeta}
\eeq
For a perfect fluid, the quantity $\zeta_a$ satisfies a
simple first-order evolution equation
\beq
\dot \zeta_a = \frac{\Theta^2}{3 \dot \rho} \Gamma_a,
\label{zeta_evolution_single}
\eeq
where
\beq
\Gamma_a \equiv D_a p
 -\frac{\dot{p}}{\dot  \rho} D_a \rho
\label{Gamma_def}
\eeq
is the nonlinear nonadiabatic pressure perturbation ($p$ is the pressure measured in the
frame defined by $u^a$, defined in Eq.~(\ref{Ptot})). For a
barotropic fluid, $\Gamma_a=0$ and $\zeta_a$ is conserved on {\em
all scales}.  The relation (\ref{zeta_evolution_single}) for
$\zeta_a$ can be seen as a generalization of the familiar
conservation law for $\zeta$, the linear curvature perturbation on
uniform energy hypersurfaces
\cite{Lyth:2004gb}. In the next sections,
we will analyze the form of $\Gamma_a$ in various physically
interesting situations.  

\smallskip

For a single scalar field, the comoving and uniform density
integrated expansion perturbations $\zeta_a$ and
$\R_a$ satisfy
\beq
\zeta_a + \R_a = - \frac{\dot \alpha}{\dot \rho} \left( D_a \rho -
\frac{\dot \rho}{\dot \phi} D_a \phi \right),
\eeq
which simply follows from their respective definitions. The right
hand side can be interpreted as the ``shift'' between
hypersurfaces of constant $\rho$ and hypersurfaces of constant
$\phi$ and the term inside the parenthesis represents the
nonlinear generalization of the so-called {\em comoving energy
density} perturbation of a single scalar field.

By choosing $u^a=u^a_{\rm com}$  as defined in Eq.~(\ref{u_comoving}), 
the energy-momentum
tensor of a single scalar field can be written in
the perfect fluid form, i.e., with vanishing $q_a$ and $\pi_{ab}$ 
and the energy density $\rho$ and
pressure $p$ given by
\beq
\rho= P_{,X}
\dot{\phi}^2-P\,, \qquad p= P
\,, \label{rho_P_single}
\eeq
as can be checked by specializing Eqs.~(\ref{rhotot}--\ref{as}) to a
single field and setting $D_a \phi=0$.

After substitution in  the definition (\ref{Gamma_def}) of
the nonadiabatic pressure covector $\Gamma_a$, we have
\beq
\Gamma_a = 
\frac{\dot{\phi}^2}{\dot{\rho}}
\,P_{,X}
\,
\left(
-
\left(1+\frac{1}{c_s^2} \right)
\,P_{,\phi} 
+P_{,X \phi}
\dot{\phi}^2
\right)
\,
 D_a \dot{\phi} 
\label{gamma_single}
\eeq
where we are using $D_a \phi=0$ (implying
also $2X\,=\,\dot{\phi}^2$). 
We also used 
\be
c_s^2 \,=\,\frac{p_{,X}}{\rho_{,X}}\,=\,\frac{P_{,X}}{P_{,X}
+\dot{\phi}^2 P_{,XX}}\,.
\ee
Then, 
 the evolution equation of $\zeta_a$ for a single
scalar field is given by
\beq
\dot \zeta_a = 
\frac{D_a \dot{\phi}}{3 \PX\,\dot{\phi}^2}
\,
\left[
-
\left(1+\frac{1}{c_s^2} \right)
\,P_{,\phi} 
+P_{,X \phi}
\dot{\phi}^2
\right]\label{expzeta}
\eeq
where we have used $\dot\rho=-\Theta\,P_{,X}\dot\phi^2$ (that can be obtain from 
the equation of motion for $\phi$)  to get this
expression.

We will show in Sec. \ref{sec:ls}, as a particular
case of a more general discussion,  that the quantity
 on the right hand side of 
equation 
(\ref{expzeta})
 vanishes on large
 scales, so that $\zeta_a$ is conserved in this limit.

\section{The two-field case}
\label{sec:two}

\subsection{Definition of adiabatic and entropy covectors}

In the two-field case, it is convenient to introduce a particular
basis in the field space in which various field dependent
quantities are decomposed into adiabatic and entropy
components. In the linear theory, this decomposition
 was first introduced in
\cite{Gordon:2000hv} for two fields. For the multi-field case, it
is discussed in \cite{bartjan} in the linear theory 
(see also \cite{DiMarco:2002eb}) and in
\cite{Rigopoulos:2005xx} in the nonlinear context.

In our case, the corresponding basis consists, in the
two-dimensional field space, of   a unit vector $\e_\sigma^I$
defined in the direction of the velocity of the two fields, and thus
{\em tangent} to the trajectory in field space, and of a  unit
vector $\e_s^I$ defined along  the direction {\em orthogonal} to
it (with respect to the field space metric), namely
\beq
\e_\sigma^\I \equiv \frac{\dot{\p}^I}{\dot{\sigma}},  \qquad G_{IJ}
\,
\e_s^I \e_s^J=1, \qquad G_{IJ}\,\e_s^I \e_\sigma^J=0\,,
 \label{e1} \label{e2}
\eeq
with
\beq
\dot{\sigma} \equiv \sqrt{ G_{IJ} \dot{\p}^I\dot{\p}^J}\,.
\eeq
Notice that generically,
the quantity $\sigmad$ {\em is not} the derivative along $u^a$ of a
scalar field $\sigma\,$; it is merely a notation.
An important consequence of the above definitions is the identity
\beq
 \delta_{\,\,\J}^\I = \e_\sigma^\I \e_\sigma{}_\J
+ \e_s^\I \e_s{}_\J \, . \label{projector}
\eeq
  From $e_{\s I} e_{\s}^I =1$, one deduces that $\Du \e_\sigma^\I$ is proportional to $e_s^I$. It is 
then convenient to define $\dot \theta$ by 
\beq
\Du \e_\sigma^\I =  \dot \theta \e_s^\I  , \qquad
\Du \e_s^\I = - \dot \theta
\e_\sigma^\I. \label{angle_1}
\eeq
Again, it is simply a short hand notation; $\dot \theta $ {\em is not} the derivative along $u^a$ of an angle $\theta$, although such an angle can be defined if the field space metric is trivial \cite{Langlois:2006vv}.

\smallskip

Making use of the basis (\ref{e1}), one can then introduce  two
linear combinations of the scalar field gradients and thus define
two covectors, respectively denoted by $\sigma_a$ and $s_a$, as
\bea
\sigma_a \tackl \equiv \tackr \e_{\sigma I} \nabla_a \p^\I
\label{tan_ort1}, \\
 s_a \tackl \equiv \tackr \e_{s I} \nabla_a \p^\I
  \label{tan_ort2}.
\eea
We will call these two covectors the {\em adiabatic} and {\em
entropy}  covectors, respectively, by analogy with the similar
definitions in the linear context \cite{Langlois:2008mn}. Whereas
the entropy covector $s_a$ is orthogonal to the four-velocity
$u^a$, i.e., $u^a s_a = 0$, this is not the case for $\sigma_a$
which contains a 'longitudinal' component:
$u^a\sigma_a=\dot\sigma$. It turns out to be useful to introduce the
spatially projected version of (\ref{tan_ort1}-\ref{tan_ort2}),
\def\sp{\sigma^{_\perp}}
\beq
\sp_a \equiv \e_{\sigma I} D_a \p^\I = \sigma_a+
\dot\sigma u_a \, , \qquad
s^\perp_a \equiv
\e_{s \I} D_a \p^\I = s_a\, .
\label{perp}
\eeq

\subsubsection{Adiabatic Klein-Gordon equation} 

We start our analysis of the evolution
equations  considering the {\it adiabatic}
 combination of the
Klein-Gordon equations, i.e., the contraction of (\ref{compact-evol-p}) by
$\e_{\sigma \I}$. By noting that we can write
\beq
\sigmad=\e_{\sigma \I} \dot \p^\I\, \qquad \ddot\sigma = \e_{\sigma \I} \Du
\dot{\p}^\I,
\eeq
and by defining \footnote{Similar notations will be employed in the following, such as $P_{,X \s} \equiv e_{\s}^I P_{,X I}$, $P_{,X s} \equiv e_{s}^I P_{,X I} \dots $ 
}
\beq
P_{, \sigma} \equiv \e_\sigma^\I P_{,I}\,,
\label{hgh}
\eeq
one obtains
\beq
 \ddot \sigma +\left(  \Theta\, + \frac{\dot{P}_{,X}}{\PX} \right) \,\dot{\sigma}-\frac{1}{\PX}P_{,\sigma}
 -\, \e_{\sigma \I} \Da \left( D^a \phi^I \right)
 -a^a\sp_a - \frac{D^b P_{,X}}{\PX}\,\sp_b=0.
\eeq
The fourth term can be rewritten as
\beq
\e_{\sigma \I} \Da \left( D^a \phi^I \right)=\Da  \left(\e_{\sigma \I}  \Dc^a
\p^\I\right)- (\Da \e_{\sigma \I} ) \Dc^a \p^\I= D^a\sp_a-(\e_{s\I}
\Da \e_\sigma^\I) s^a,
\eeq
where we have used the definition of $\sigma_a^\perp$, Eq.~(\ref{perp}),
 and the identity (\ref{projector}). Inserting
 \beq
 \mathcal{D}_a \dot \p^I= u^b \mathcal{D}_b \left( \nabla_a \p^I \right)+ \left( \nabla_a u^b \right) \left( \nabla_b \p^I \right)
 \label{identity} 
 \eeq
into the calculation of the Lie derivative of $s_a$ with respect to $u_a$, one gets
\beq
\e_{s \I} \mathcal{D}_a \e_\sigma^\I=-\e_{\s \I} \mathcal{D}_a \e_s^\I=
\frac{1}{\dot \sigma} (\dot s_a + \dot \theta \sigma_a )
\, ,
\label{esDesigma}
\eeq
from which one obtains
\beq
\e_{\sigma \I} \Da \left( D^a \phi^I \right)=D^a \sp_a- \frac{ 1}{\dot \sigma} (\dot s_a + \dot \theta \sigma_a^\perp )s^a\,.
\eeq
The adiabatic combination of the Klein-Gordon
equations can thus be written as
\beq
 \ddot \sigma + \left( \Theta\, +\frac{ \dot{P}_{,X}}{ P_{,X}} \right)\,\dot{\sigma}-\frac{P_{,\sigma}}{ P_{,X}}
=\, \nabla^a \sp_a
  -Y_{(s)} \,,
\label{evolsigma-non-explicit}
\eeq
where we have defined
\beq
Y_{(s)} \equiv \frac{ 1}{\dot \sigma} (\dot s_a + \dot \theta \sigma_a^\perp )s^a-\frac{D^a P_{,X}}{P_{,X}}\sp_a\, \,,
\label{Y_s}
\eeq
and used the property
\beq
D^a\sp_a+a^a\sp_a=\nabla^a\sp_a, \label{grad}
\eeq
which is valid for any covector {\it orthogonal to} $u^a$ 
(and thus also for $s_a$). Note that our $Y_{(s)} $ coincides with the one of \cite{Langlois:2006vv} in the case of canonical kinetic terms, in which situation the last term in (\ref{Y_s}) vanishes. \\
Second derivatives of the scalar fields are hidden in the term $\dot{P}_{,X}$. Explicitely, this gives
\beq
\frac{\ddot \s}{c_s^2}+\Theta \dot \s+\frac{P_{,X \s}}{\PX}\dot \s^2-\frac{P_{,\s}}{\PX}=\nabla^a \sp_a
  -Y_{(s)}  +\frac{\dot \s P_{,XX}}{2 \PX}  \Tdot{(\s^{\perp}_a  \sigma^{\perp\,a}+s_a s^a)}\,,
  \label{evolsigma}
\eeq
where we have introduced the important, \textit{spacetime dependent}, quantity

\beq
c_s^2 \equiv \frac{\PX}{\PX+ \dot \s^2 P_{,XX}}\,.
\eeq
Notice
that $c_s$ is a fully \textit{nonlinear} quantity; this implies
for example 
that  one cannot replace $\dot \s^2$ by $2X$ in its
definition, since these quantities  differ by terms in spatial gradients (see (\ref{Xdecomposition})). At linear order however, $c_s$ coincides with the now well known speed of sound of \textit{k}-inflationary models \cite{ArmendarizPicon:1999rj,Langlois:2008mn}.

\subsubsection{Entropic Klein-Gordon equation}
 
Let us now consider the {\it entropic} combination of the
Klein-Gordon equations, i.e., the contraction of (\ref{evolp})
with $\e_{s \I} $. By using
\beq
\e_{s \I} \Du \dot \p^\I=\dot\theta\dot\sigma,
\eeq
and by defining the entropic gradient of the potential,
\beq
P_{, s} \equiv \e_s^\I P_{,I},
\eeq
one finds
\beq
P_{,X}\,\dot \sigma\dot\theta -P_{,s} -P_{,X}\,
\e_{s \I} \Da \left( D^a \phi^I \right)-
P_{,X}\,
a^a
s_a-D^b P_{,X}\,s_b=0.
\eeq
One can rewrite the last three terms by using the identity
\beq
\e_{s \I} \Da \left( D^a \phi^I \right)=D^a s_a+ \frac{1}{\dot \sigma} (\dot s_a + \dot \theta
\sigma^\perp_a )\sp{}^a,
\eeq
derived by inserting (\ref{identity}) into the calculation of the Lie derivative of $\s_a$ with respect to $u_a$. Applying
the property (\ref{grad}) to the  covector $s_a$, one finally gets
\beq
\dot \sigma \dot \theta - \frac{P_{,s}}{ P_{,X}}\,
= \,\nabla_a s^a
  + Y_{(\sigma)} \,,\label{thetadot}
\eeq
with
\beq
Y_{(\sigma)}\equiv \frac{1}{\dot \sigma} (\dot s_a + \dot \theta
\sigma^\perp_a )\sp{}^a+\frac{D^a P_{,X}}{ P_{,X}} s_a\,.
\label{Y_sigma}
\eeq

\bigskip

To conclude, we have shown how to replace the Klein-Gordon
equations for the fields by two new equations
describing the evolution along the adiabatic and
entropy directions respectively.  Our covariant equations (\ref{evolsigma}) and
(\ref{thetadot}) look very similar to the homogeneous equations:
on the other hand 
they capture the fully nonlinear dynamics of the scalar fields.
 Whereas these equations generalize the {\em
background} evolution equations,  we will proceed in
the next subsection to derive fully nonlinear and exact
equations for the covectors, 
which mimic and generalize the {\em linearized}
equations for the adiabatic and entropy components.

\subsection{Evolution of the  covectors}

We now derive evolution equations for the covectors $\sigma_a$ and
$s_a$. More precisely, our purpose is  to find two evolution
equations, which are second order in time (with respect to
the Lie derivative along $u^a$) and  which mimic the equations
obtained in the linear theory for the perturbations
$\delta\sigma$ and $\delta s$ (see \cite{Langlois:2008mn}).

\subsubsection{
 Evolution equation for
the adiabatic covector}
  
Starting from the definition of $\sigma_a$ in  (\ref{tan_ort1}),
 one finds that its  time derivative, i.e., the
Lie derivative with respect to $u^a$, is given by
\beq
\dot\sigma_a=\e_{\sigma \I} \mathcal{D}_{a}  \dot\p^\I+\dot\theta s_a=\nabla_a\dot\sigma+\dot\theta s_a,
\label{dotsigma}
\eeq
where the last equality is obtained by using $\dot\p^\I=\dot\sigma
\e_{\sigma}^ \I$. A further time derivative yields
\beq
\ddot \sigma_a =\nabla_a\ddot\sigma+ \ddot \theta s_a
+  \dot \theta \dot s_a.
\eeq
The next step consists in using (\ref{evolsigma-non-explicit})
to eliminate $\ddot\sigma$ in the  above expression.
This gives

\bea
&&  \ddot \sigma_a +  \Theta \dot \sigma_a +\dot\sigma  \nabla_a \left( \Theta + \frac{\dot{P}_{,X}}{\PX} \right)
+ \frac{\dot{P}_{, X}\,\dot{\sigma}_a}{\PX}
+\frac{P_{,\sigma}\,\nabla_a P_{,  X} }{P_{,  X}^2}
-\left(\frac{P_{,\sigma
\sigma}}{\PX}+\dot \theta \frac{P_{,s}}{{\PX}\,\dot \sigma} \right) \sigma_a 
- \nabla_a \left(\nabla^c \sp_c\right)
 - \frac{P_{, X \sigma} \,\nabla_a X}{\PX} 
 \nonumber \\
&& \hskip 0.3 cm
\,=\,\left( \dot \theta+\frac{P_{,s}}{\dot \sigma \PX}\right) \dot s_a  +\left[\ddot\theta +\frac{P_{,\sigma s}}{\PX}+ \dot \theta\, \left(\Theta+\frac{\dot \PX}{\PX}\right)\right]\, s_a - \nabla_aY_{(s)}\ , \label{sigma1}
\eea
where we have used the relation
\beq
\nabla_a P_{,\sigma}= P_{, \sigma \sigma} \sigma_a +P_{,\sigma s} s_a
+\frac{P_{,s}}{\dot \sigma} (\dot s_a + \dot \theta \sigma_a)
+P_{, X \sigma} \,\nabla_a X \,,
\eeq
and introduced the notation
\beq
P_{,\sigma \sigma} \equiv \e_\sigma^\I \e_\sigma^\J \mathcal D_I \mathcal D_J P, \qquad P_{,s s} \equiv \e_s^\I \e_s^\J \mathcal D_I \mathcal D_J P,
\qquad P_{,s \sigma} \equiv \e_s^\I \e_\sigma^\J \mathcal D_I \mathcal D_J P\,\,\,.
\eeq
In the previous expressions,
$ \mathcal D_I$ denotes the covariant derivative associated with $G_{IJ}$ (thus having $ \mathcal D_I  \mathcal D_J P \equiv P_{,IJ}-\Gamma_{IJ}^K P_{,K}$). 

The evolution equation for $\sigma_a$ can then be  decomposed
into a longitudinal part, obtained by
contracting (\ref{sigma1}) with  $u^a$, and an orthogonal
part obtained by contraction with
$h_{ab}$. By  using the relation
\beq
\dot P_{,\sigma} = P_{, \sigma \sigma} \dot \sigma + \dot \theta P_{,s}
+P_{, X \sigma} \, \dot{X}
,
\eeq
it is not difficult to see that the longitudinal part yields in
fact the time derivative of (\ref{evolsigma}). What is more
interesting is the orthogonal or spatial part.  Using several equations collected in the appendix, one can expand the various quantities inside the previous expression in terms of the perturbations
$\sigma_a^\perp$ and $s_a$. After a long but straightforward calculation, the evolution equation for the adiabatic covector can be written in a form close its linear counterpart \cite{Langlois:2008mn}:

\bea
&&\left( \ddot \s_a  \right)^{\perp} +\left[ \Theta+  \frac{c_s^2}{\PX} \Tdot{\left(\frac{\PX}{c_s^2}\right)} +(c_s^2-1)\frac{\W}{\dot \s} \right] \, \left( \dot \s_a  \right)^{\perp}
+\frac{c_s^2}{\PX}\left[-P_{,\s\s}-\frac{\dot \theta P_{,s}}{\dot \s} +e^{-3 \alpha} \Tdot{\left(  e^{3\alpha} \dot \s P_{,X\s}\right)}   -P_{,X \s} \W \right]\s_a^{\perp}
\nonumber \\
 &&\hskip0.5cm + \dot \s c_s^2 D_a \Theta-c_s^2 D_a \left(\nabla^c \s_c^{\perp}  \right)=
 \left[\Xi +\frac{\Z}{\dot \s}\right]\dot s_a +\left[\dot \Xi-\Xi\left(  2\frac{\dot c_s}{c_s} -\frac{P_{,\s}}{\PX \dot \s}     \right)  +M\right] s_a
 \nonumber \\
 &&\hskip0.5cm -c_s^2 D_a Y_{(s)}
 +\frac{1}{2 \dot \s}(1-c_s^2)D_a \dot \Pi
+\frac{c_s^2}{2 \PX}\left[P_{,X \s}-P_{,XX} \W+e^{-3 \alpha} \Tdot{\left( e^{3 \alpha} \dot \s P_{,XX} \right)}\right] D_a \Pi
 \label{sigma-similar-linear}
\eea
In the previous equation we introduced various
quantities:
\bea
 \Z&\equiv& \nabla_a s^a
  + Y_{(\sigma)}\, ,\qquad \qquad \W\, \equiv\, \nabla^a \sp_a-Y_{(s)}\,, \label{def-Z} \\  
  \Pi &\equiv& \sp_c
{\sp}^c +s_c s^c\,, \label{def-Pi}  \\ 
\Xi&\equiv& \frac{1}{\dot \s \PX}\left[ (1+c_s^2) P_{,s}-c_s^2 \dot \s^2P_{,Xs}\right]\,,
\label{def-Xi}
\eea
and 
\beq
M= \frac{\dot \Z}{\dot \s}+\frac{c_s^2 P_{,s} \W}{\PX \dot \s^2} +\frac{ c_s^2 \Z}{\dot \s}\left(2 \Theta+\frac{P_{,X\s} \dot \s}{\PX}+\frac{\dot \s P_{\s} P_{,XX}}{\PX^2}-2\frac{\dot c_s}{c_s^3}-\frac{P_{,XX} \dot \Pi}{\PX}-\frac{\W}{\dot \s}  \right) +\frac{c_s^2\,  \dot \Pi}{2 \PX \dot \s}\left[ P_{,X s}-\frac{P_{,XX}}{\PX}P_{,s}  \right] \,.\nonumber \\
\eeq
Moreover, we used the integrated volume expansion $\alpha$
introduced in Eq.~(\ref{alpha_def}).\\

Let us make a few comparisions with the case of a standard Lagrangian: the friction term is not only given by the volume expansion $\Theta$ but has a non vanishing contribution coming from the nonlinear speed of sound $c_s$, which may vary in time $\left(\dot c_s \neq 0 \right)$ and may be different from unity $\left(c_s^2 \neq 1\right)$. The last two terms in (\ref{sigma-similar-linear}), which exhibit a rich dependence on the form of the kinetic term, also vanish for a standard Langrangian $P=X-V(\p^I)$. Moreover, in that case, $\Xi$ reduces to $2 \dot \theta$, which encodes all the multi-field effects at linear order, as shown in  \cite{Gordon:2000hv}. The important parameter $\Xi$ plays an equivalent role for Lagrangians of the form $P(X,\p^I)$ studied in this paper, as shown in \cite{Langlois:2008mn}: if $\Xi=0$, the situation is effectively single-field at linear order.

\subsubsection{
 Evolution equation for
the entropy covector}  

Let us now consider the evolution equation for $s_a$. From
Eq.~(\ref{esDesigma}) the time derivative of $s_a$ is given by
\beq
\dot s_a= \dot\sigma \, \e_ {s\I} \mathcal{D}_a \e_{\sigma}^\I -\dot\theta
\sigma_a.
\eeq
Taking another time derivative, one finds
\beq
\ddot s_a = - \ddot \theta \sigma_a -  \dot \theta
\dot \sigma_a+\frac{\ddot\sigma}{\dot\sigma}(\dot s_a + \dot \theta \sigma_a )+\dot\sigma\nabla_a \dot\theta-\dot \s^2 R^I_{KLJ}e_{s I}e_{\s}^J e_{\s}^K\nabla_a \p^L,
\eeq
where we have used (\ref{esDesigma}) and (\ref{angle_1}) and the equality
\beq
\mathcal D_a\left( \Du A^I\right)-\Du \left( \mathcal D_a A^I \right)=  R^I_{KLJ} \nabla_a \p^L \dot{\p}^J A^K
\eeq
applied to the vector $e_{\s}^I$ and where $R^I_{KLJ}$ is the Riemann tensor associated to the metric 
$G_{IJ}$\footnote{We thus have $R^I_{KLJ}=\partial_L \Gamma^I_{KJ}-\partial_J \Gamma^I_{KL}+\Gamma^I_{LM} \Gamma^M_{KJ}- \Gamma^I_{JM} \Gamma^M_{KL}.$}. We can
now use the entropic equation (\ref{thetadot}) to get rid of
$\nabla_a \dot\theta$. Furthermore, using the relation
\beq
\nabla_a P_{,s}= P_{, s \sigma} \sigma_a +P_{,ss} s_a
-\frac{P_{,\sigma}}{\dot \sigma} (\dot s_a + \dot \theta
\sigma_a)+ P_{, X s} \,\nabla_a X,
\eeq
we obtain 
\bea
&&\ddot s_a - \frac{1}{\dot \sigma}\left(\ddot \sigma - \frac{P_{,\sigma}}{\PX}\right)
\dot s_a -\left( \frac{P_{,ss}}{\PX}+\dot \theta^2 \right) s_a
-\nabla_a \left(\nabla_c s^c\right) -\frac{P_{, X s}}{\PX} \,\nabla_a X
+\dot \s^2 R^I_{KLJ}e_{s I}e_{\s}^J e_{\s}^K\nabla_a \p^L
 \nonumber \\
&& \qquad
=- 2
\dot \theta \dot \sigma_a  -P_{,s}\,\frac{\nabla_a P_{,X}}{ P_{,X}^2} 
 +\left[\frac{\dot
\theta}{\dot \sigma} (\ddot \sigma-\frac{P_{,\sigma}}{\PX})
-\ddot \theta +\frac{P_{,\sigma s}}{\PX}\right] \sigma_a+
\nabla_a Y_{(\sigma)}\ .\label{s0}
\eea
As for the adiabatic equation,
 the longitudinal part of this equation,
upon using the relation
\beq
\dot P_{,s}  =  P_{,\sigma s}\dot \sigma - \dot \theta P_{,
\sigma} +P_{,Xs}\,\dot{X} , \label{dotVs}
\eeq
yields the time derivative of
Eq.~(\ref{thetadot}). The orthogonal part, instead, yields
\bea
&&\ddot s_a - \frac{1}{\dot \sigma}\left(\ddot \sigma - \frac{P_{,\sigma}}{\PX}\right)
\dot s_a -\left( \frac{P_{,ss}}{\PX}+\dot \theta^2-\dot \s^2 R_{IKLJ}e_{s}^{ I}e_s^Le_{\s}^J e_{\s}^K  \right) s_a
-D_a \left(\nabla_c s^c\right) -\frac{P_{, X s}}{\PX} \,D_a X
 \nonumber \\
&& \qquad
=- 2
\dot \theta \left( \dot \sigma_a\right)^{\perp}  -P_{,s}\,\frac{D_a P_{,X}}{ P_{,X}^2} 
 +\left[\frac{\dot
\theta}{\dot \sigma} (\ddot \sigma-\frac{P_{,\sigma}}{\PX})
-\ddot \theta +\frac{P_{,\sigma s}}{\PX}\right] \left(\sigma_a \right)^{\perp}+
D_a Y_{(\sigma)}\,,\label{s1}
\eea
where we have used  the property that the covectors $\dot s_a$ and
$\ddot s_a$ are purely spatial, i.e., that $(\dot s_a)^\perp =
\dot s_a $ and $(\ddot s_a)^\perp = \ddot s_a $. 
Using (\ref{thetadot}) as well as
 various expressions collected in the appendix, we finally obtain
\bea
&&\ddot s_a+ \left( \Theta+\frac{\dot{P}_{X}}{\PX}-\frac{\W}{\dot \s} \right)
\dot s_a +\left(\mu_s^2-\frac{\Z^2}{\dot \s^2}-\frac{\Z \Xi}{\dot \s c_s^2}\right) s_a-D_a \left(\nabla_c s^c\right) 
 \nonumber \\
&& \qquad
=- 
\left(\frac{\Xi}{c_s^2}+2\frac{\Z}{\dot \s}\right) \left( \dot \sigma_a\right)^{\perp}  
 +\left[  \frac{\ddot \s }{\dot \s}\left(\frac{\Xi}{c_s^2}+2\frac{\Z}{\dot \s} \right)-\frac{\dot \Z}{\dot \s}\   \right] \left(\sigma_a \right)^{\perp}+
D_a Y_{(\sigma)}
 \nonumber \\
&& \qquad \hskip0.4cm
-\frac{1}{2\PX}\left(P_{,Xs}-\frac{P_{,XX} P_{,s}}{\PX} \right) \left(D_a \Pi-\frac{\dot \Pi}{\dot \s}\sigma_a^{\perp}   \right)\,,\label{s2}
\eea
with 
\bea
\mu_s^2 \equiv
- \frac{P_{,ss}}{\PX}+\dot \s^2 R_{IKLJ}e_{s}^{I}e_s^Le_{\s}^J e_{\s}^K -\frac{P_{,s}^2}{c_s^2 \dot \s^2 \PX^2}+\frac{2 P_{,Xs} P_{,s}}{\PX^2} \,
\label{mus}, 
\eea
which mimics its linear counterpart in \cite{Langlois:2008mn}. Let us remark that the last term in (\ref{s2}), absent in the linear theory, also vanishes for a standard Lagrangian.

\smallskip

Starting from the fully nonlinear Klein-Gordon equations, we
thus  obtained a system of two coupled equations
(\ref{sigma-similar-linear}) and (\ref{s2}), controlling
 the evolution
of our nonlinear adiabatic and entropy components. 
Although they appear quite involved, we will learn that
they simplify considerably in physically interesting
situations, in particular when considering linear or  large
scale limits. Moreover, 
 as we will see later, immediately deduced from
these equations are the already known evolution  equations for the
{\em linear} adiabatic and entropy components. Furthermore, since
our equations are exact, they can be used  to go beyond the linear
order,  up to second or higher orders in the expansion.

\subsection{Generalized covariant perturbations}

In this subsection, we will be first interested in the covariant
generalization of the comoving energy density and curvature
perturbations in the context of a two-field system. We will then
consider the generalization of the curvature perturbation on
uniform energy density hypersurfaces.

\subsubsection{Comoving energy density covector}

Let us  introduce the covector
\beq
\epsilon_a\equiv\Dc_a\rho- \frac{\dot \rho}{\dot \sigma}\sp_a,
\label{epsilon}
\eeq
which can be interpreted as a covariant generalization of the {\em comoving
energy density} perturbation. In order to obtain  the explicit expression of
$\epsilon_a$ in terms of $\sigma_a$ and $s_a$, let us rewrite the components 
(\ref{rhotot}-\ref{as})
of the energy-momentum tensor in the form
\bea
\rho \tackl = \tackr P_{,X}    \dot \sigma^2 - P ,
\label{EMTrho} \\
\pr
 \tackl = \tackr \frac{P_{,X}}{3} \Pi +P ,
\label{EMTP} \\
q_a \tackl = \tackr - P_{,X}\dot\sigma \sp_a, \label{qa} \\
\pi_{ab} \tackl = \tackr  P_{,X}
\left(  \Pi_{ab} -\frac13 h_{ab} \Pi \label{pi}
\right),
\eea
where we have defined
\beq
\Pi_{ab} \equiv \sp_a \sp_b +s_a s_b\, ,
\eeq
while the quantity
 $\Pi$ has already been introduced in (\ref{def-Pi}).\\
Using (\ref{EMTrho}), one finds
\bea
D_a \rho  &=&
\left(P_{,XX} \dot{\sigma}^2-P_{,X}\right)
\left( \dot{\sigma} D_a \dot{\sigma}-\frac{D_a \Pi}{2}
\right)+2 P_{,X} \dot{\sigma} D_a \dot{\sigma}-P_{,\sigma} \s_a^{\perp}-P_{,s} s_a
+\dot{\sigma}^2\ \left( P_{,X \sigma}\s_a^{\perp}+P_{,Xs} s_a\right)\,, \nonumber
 \\
\dot{\rho} &=&
\left(P_{,XX} \dot{\sigma}^2-P_{,X}\right)
\left( \dot{\sigma} \ddot{\sigma}-\frac{\dot{\Pi}}{2}
\right)+2 P_{,X} \dot{\sigma} \ddot{\sigma}-P_{,\sigma} \dot{\sigma}
+P_{,X \sigma}\,\dot{\sigma}^3 \,,
\eea
 from which one obtains
\beq
\epsilon_a=\frac{\PX}{c_s^2}\left( \dot \s \left( \dot \s_a \right)^{\perp}-\ddot \s \s_a^{\perp} \right)-\frac{\PX}{c_s^2}\left( \dot \s \Xi+\Z \right) s_a+\frac 1 2 (\PX-\dot \s^2 P_{,XX}) \left(D_a \Pi-\frac{\dot \Pi}{\dot \s}\s_a^{\perp} \right)\,.
\label{epsilon_a}
\eeq

This expression can be employed to rewrite Eq. (\ref{s2}) for the entropy covector $s_a$ in a useful different form. As noticed above, the longitudinal projection of Eq. (\ref{s0}) yields the time derivative of (\ref{thetadot}), which reads

\beq
-\frac{\dot \theta}{\dot \s}\frac{P_{,\s}}{\PX}-\ddot{\theta} +\frac{P_{,\s s}}{\PX}=\frac{\dot \theta}{\dot \s}\ddot \s-\frac{P_{,X s} \dot X}{\PX \dot \s}+\frac{\dot \PX P_{,s}}{\dot \s \PX^2}-\frac{\dot Z_{(\s)}}{\dot \s}\,.
\eeq
This relation, together with Eqs. (\ref{evolsigma-non-explicit}) and (\ref{epsilon_a}), enables us to reexpress (\ref{s2}) as

\bea
&& \ddot s_a+\left[\Theta+\frac{\dot \PX}{\PX}-\frac{\W}{\dot \s}\right]\dot s_a+\left[\mu_s^2+\frac{\Xi^2}{c_s^2}+\frac{\Z}{\dot \s}\left( \frac{\Z}{\dot \s}+2\, \Xi \right) \right] s_a-D_a \left( \nabla_c s^c \right)
 \nonumber \\
&& \qquad=
-\frac{1}{\PX \dot \s}\left( \Xi+2\frac{c_s^2}{\dot \s}\Z \right) \epsilon_a+\frac{1}{\dot \s}\left( 1-\frac{1}{2 c_s^2} \right) \left( \Xi+2\frac{c_s^2}{\dot \s}\Z \right)\left( D_a \Pi -\frac{\dot \Pi}{\dot \s}\s_a^{\perp} \right)
\nonumber \\
&& \qquad-\frac{\dot \Z}{\dot \s}\s_a^{\perp}+D_a Y_{(\s)}\,.
\label{s3}
\eea

 As in the linear theory \cite{Langlois:2008mn}, this gives us an alternative expression for the evolution equation for $s_a$, in which the comoving energy density perturbation appears explicitely on the right hand side. This expression will be useful in Sec.~\ref{sec:ls} when discussing the large-scale evolution of $s_a$.
 
\subsubsection{Comoving curvature covector}

Together with the comoving energy density, 
the comoving curvature perturbation can be generalized. For the general case of several
scalar fields,  this is done by defining the comoving
integrated expansion perturbation (recall the expression for $q_a$ in (\ref{qa})), 
\beq
\R_a \equiv   - D_a \alpha  -  \frac{\dot \alpha}{\PX \dot \sigma^2} q_a. \label{R_N}
\eeq
The definition of the Sasaki-Mukhanov covector given for a single
scalar field in Eq.~(\ref{Q_single}) can then be extended to the
case of several fields, by defining for each field
\beq
Q^I_a \equiv D_a \phi^I - \frac{\dot \phi^I}{\dot \alpha}
D_a \alpha\,.
\eeq
Thus, the comoving covector $\R_a$ can also be written as
\beq
\R_a = \frac{\dot \alpha}{(\dot \phi_J \dot \phi^J)}\  \dot
\phi_I Q^I_a .
\eeq
In the {\em two}-field case, the definition (\ref{R_N}) reduces,
using (\ref{qa}),  to
\beq
\R_a \equiv - D_a \alpha + \frac{\,\dot \alpha}{\dot \sigma}
\sigma_a^\perp. \label{R_two}
\eeq
Furthermore, one can generalize the Sasaki-Mukhanov variable to
the {\em adiabatic} covector by defining
\beq
Q_a \equiv\e_{\sigma \I} Q^I_a = \sigma_a^\perp - \frac{\dot \sigma}{\dot \alpha} D_a
\alpha . \label{Q_two}
\eeq
Following \cite{Langlois:2006vv}, instead of giving the evolution equation of $\R_a$, we will use
 a fluid description by considering the covariant
generalization of the uniform density curvature perturbation,
i.e., the integrated expansion perturbation on
uniform density hypersurfaces  $\zeta_a$, defined
in Eq.~(\ref{zeta}). In the two-field case, these two quantities
are related by
\beq
\zeta_a + \R_a = - \frac{\dot \alpha\,}{\dot \rho} \epsilon_a.
\label{zeta_R_two}
\eeq

In contrast with the case of a single scalar field, which can
always be described as a perfect fluid, the total energy-momentum
for two or more scalar fields corresponds  in general to that
of a dissipative fluid: the nonlinear formalism developed in
\cite{Langlois:2006iq} will thus be useful in this case.

The adiabatic Klein-Gordon equation (\ref{evolsigma}) can be rewritten as
a continuity equation for the total energy density (\ref{EMTrho}) and pressure
(\ref{EMTP}), which reads
\beq
\label{conservation}
\dot \rho + \Theta (\rho +\pr) = {\cal D},
\eeq
with the {\it dissipative} term 
\beq
\mathcal{D}=\PX \left( \dot \s\W +\frac{1}{3} \Theta \Pi +\frac{1}{2}  \dot \Pi \right)\,.
\label{dissipative}
\eeq

In \cite{Langlois:2006iq} it was shown that the evolution equation
for $\zeta_a$ for a dissipative fluid, which generalizes (\ref{zeta_evolution_single}),  is
given by
\beq
\dot \zeta_a = \frac{\Theta^2}{3\dot \rho} \left(\Gamma_a +
\Sigma_a \right),
\label{zeta_evolution}
\eeq
where the second source term on the right hand side, due to the dissipative nature
of the fluid,
is defined in terms of ${\cal D}$ as
\beq
\Sigma_a \equiv - \frac{1}{\Theta} \left(D_a \D - \frac{\dot \D}{\dot
\rho} D_a \rho\right) + \frac{\D}{ \Theta^2} \left(D_a \Theta -
\frac{\dot \Theta}{\dot \rho} D_a \rho \right) . \label{Sigma}
\label{Sigma_def}
\eeq

Let us return to the covector 
 $\Gamma_a$. Substituting the expression (\ref{EMTP}) for $\pr$ into its definition (\ref{Gamma_def}), as well as using
 \bea
 \dot P &=& P_{,\s} \dot \s +\PX \left(\dot \s \ddot \s -\frac{1}{2} \dot \Pi \right)\,,
 \label{dot P}\\ 
 D_a P&=&P_{,\s} \s_a^{\perp}+P_{,s} s_a+\PX \left( \dot \s ( \dot{\sigma}_a^ \perp-\dot \theta s_a)-\frac{1}{2}D_a \Pi \right)\,,
 \eea
one obtains
\be
\Gamma_a=-\epsilon_a \frac{\dot \s}{\dot \rho}\left[ (1+c_s^2)P_{,\s}-c_s^2 \dot \s^2P_{,X \s} \right]+\PX \dot \s \Xi s_a-\PX c_s^2\left(D_a \Pi-\frac{\dot \Pi}{\dot \rho}D_a \rho \right)+\frac{1}{3}\left(D_a \left( \PX \Pi\right)-\frac{\Tdot{( \PX \Pi)}}{\dot \rho}D_a \rho \right)
 \label{Gamma_1}
\eeq
where we have introduced  $\epsilon_a$ defined in
Eq.~(\ref{epsilon}). 
The above equation expresses the
nonlinear nonadiabatic pressure perturbation in the two-field
case, which sources Eq.~(\ref{zeta_evolution}). It will be useful
in the next section, where we will consider both the linear and
super-Hubble approximations of our evolution equations.

The evolution of $\zeta_a$,
governed by Eq.~(\ref{zeta_evolution}), is thus sourced by a
rather complicated term obtained by summing $\Gamma_a$ 
 and $\Sigma_a$ in Eq.~(\ref{Sigma_def}). We
will learn in the next sections that this term simplifies considerably
in physically interesting cases, 
when either the linear or the super-Hubble approximations are
taken.

\section{Approximate equations}
\label{sec:approximate}

 In this section we study the evolution equations of the
two-field system under two types of approximations: the linear
limit and the limit in which we neglect higher orders in spatial
gradients. In an expanding FLRW universe, the latter corresponds
to the {\em large scale} limit. In the rest of the paper we will
use the symbol $\simeq $ to denote an equality at the linear level,
and the symbol $\approx $ to denote an equality valid only on large
scales.

\subsection{Homogeneous and linearized equations}

In many cosmological applications, since our Universe appears to
be close to a FLRW universe on large scales, it is sufficient to
restrict oneself to  the {\it linearized} version of the evolution
equations. We first consider this linearization procedure
directly at the level of  the covariant equations, as done in
\cite{Ellis:1989jt}.

In a strictly FLRW universe, all the spatial gradients vanish and
therefore
\beq
\sigma_a^\perp=0\qquad, \qquad s_a=0\,\,.\qquad ({\rm FLRW})
\eeq
Consequently, the scalar quantities $Y_{(s)}$ and $Y_{(\sigma)}$,
defined respectively in (\ref{Y_s}) and (\ref{Y_sigma}),  vanish
and the evolution equations for $\sigma$ and $\theta$,
respectively (\ref{evolsigma-non-explicit}) and (\ref{thetadot}), reduce to
\bea
 \ddot \sigma + \left( 3H   +\frac{ \dot{P}_{,X}}{ P_{,X}} \right) \,  \dot \sigma 
 -\frac{P_{,\sigma}}{ P_{,X}}
&=&0\,, \qquad ({\rm FLRW})
\label{evolsigma_RW}
\\ 
\dot \sigma \dot \theta - \frac{P_{,s}}{ P_{,X}}\,
& = &0
\qquad ({\rm FLRW})
 \label{thetadot_RW}
\eea
where we have introduced the Hubble parameter $H= \Theta/3$. Not surprisingly,
the above equations exactly correspond to the homogeneous equations
given in \cite{Langlois:2008mn}. Furthermore,
in  a FLRW universe, all the terms in Eq.~(\ref{sigma-similar-linear}) for $\sigma_a^{\perp}$ and Eq.~(\ref{s3}) for $s_a$  vanish.

At linearized order, we treat  the covectors $\sigma_a^\perp$ and
$s_a$, vanishing at zeroth order, as {\it first-order}
quantities. Similarly, their derivatives $\dot s_a$, $\ddot s_a$,
$(\dot \sigma_a)^\perp$ and $(\ddot\sigma_a)^\perp$ are
first-order quantities. Therefore the linearized version of the
evolution equations are simply obtained by keeping only the
homogeneous terms in the coefficients multiplying the spatial
projection of $\sigma_a$, $s_a$ and their derivatives. 
The
linearized versions of (\ref{sigma-similar-linear}) and (\ref{s3}) are
thus (recall the definition of $\Xi$ in (\ref{def-Xi}))

\bea
&&\left( \ddot \s_a  \right)^{\perp} +\left[3 H+  \frac{c_s^2}{\PX} \Tdot{\left(\frac{\PX}{c_s^2}\right)} \right]\left( \dot \s_a  \right)^{\perp}
+\frac{c_s^2}{\PX}\left[-P_{,\s\s}-\PX \dot \theta^2 +e^{-3 \alpha} \Tdot{\left(  e^{3\alpha} \dot \s P_{,X\s}\right)}    \right]\s_a^{\perp}
\nonumber \\
 &&+ \dot \s c_s^2 D_a \Theta-c_s^2 D_a \left(D^c \s_c^{\perp}  \right) \simeq
\Xi \dot s_a +\left[\dot \Xi-\Xi\left(  \frac{\Tdot{(H c_s^2)}}{H c_s^2} -\frac{P_{,\s}}{\PX \dot \s}-\frac{\dot H}{H}       \right)  \right] s_a
  \label{sigma1_spatial_lin}
\eea
and
\bea
&& \ddot s_a+\left(3 H+\frac{\dot \PX}{\PX} \right)\dot s_a+\left(\mu_s^2+\frac{\Xi^2}{c_s^2} \right) s_a-D_a \left( D_c s^c \right)
\simeq
-\frac{\Xi}{\PX \dot \s} \epsilon_a
\,.\label{s1_lin}
\eea
Note that the terms involving $Y_{(s)}$, $Y_{(\sigma)}$ and $\Pi$
have disappeared, since these scalars are {\it quadratic} in
first-order quantities. 
We have also replaced $\nabla^c \sp_c$ by
$D^c \sp_c$, as well as $\nabla^cs_c$ by $D^cs_c$, since their
difference is quadratic in first-order quantities, according to
(\ref{grad}). Indeed,  the acceleration vector $a^b$, which
vanishes at zeroth order, is considered as a first-order quantity.

One can also linearize the evolution equation for $\zeta_a$.
As discussed above, the terms containing
$Y_{(s)}$ and $\Pi$ can be neglected.
The dissipative term ${\cal D}$  thus reduces to

\beq
{\cal D}\simeq \PX \dot \sigma D^a \sigma_a^\perp,
\eeq
while the expression for $\epsilon_a$ becomes

\be
\epsilon_a \,\simeq\, \frac{P_{,X}}{c_s^2}\,\left[
 \dot{\sigma} \dot{\sigma}_a^\perp-\ddot{\sigma} \sigma_a^\perp
-\dot{\sigma} \Xi s_a
\right] \,.
\ee
On the other hand,  the expression for
$\Gamma_a$ becomes

\bea
\Gamma_a &\simeq&  -\epsilon_a \frac{\dot \s}{\dot \rho}\left[ (1+c_s^2)P_{,\s}-c_s^2 \dot \s^2P_{,X \s} \right]+\PX \dot \s \Xi s_a
\,,
\eea
from which it is immediate to obtain the expression for
$\dot{\zeta}_a$.

\subsection{Expansion in spatial gradients}
\label{sec:ls}
Apart from the linearization procedure, there is another
approximation in  the cosmological context that is used to
describe the Universe on very large scales. This approximation is based on an
expansion in spatial gradients, which are small for scales larger
than the local Hubble radius
\cite{Salopek:1990jq,Comer:1994np,Deruelle:1994iz}. 
 Recently for instance, it has been adopted in \cite{Takamizu:2008ra}
  for studying nonlinear perturbations by means of 
  the ADM formalism, in a system  
  describing a single scalar field with arbitrary kinetic
  terms, up to second order in the gradient expansion. In the present article, we restrict ourselves to the leading order:
  in this perspective one sees, from their definition (\ref{perp}),
that $\sigma_a^\perp$ and $s_a$ are first-order quantities with
respect to spatial gradients because they are linear combinations
of spatial gradients. The scalars $Y_{(s)}$ and $Y_{(\sigma)}$ however are
second-order with respect to spatial gradients since they are
quadratic in $\sigma_a^\perp$ and $s_a$ (or their time
derivatives). Hence, the right hand side of Eq.~(\ref{evolsigma-non-explicit})
and of Eq.~(\ref{thetadot}) can be neglected on large scales, so
that these two equations become, in the large-scale limit,
\bea
&&
\ddot \sigma +\left( \Theta+\frac{\dot \PX}{\PX} \right) \dot \sigma -\frac{P_{,\sigma}}{\PX} \approx 0,
\label{sigma_evol_ls} \\
&& \dot \theta  -  \frac{P_{,s}}{\PX \dot \sigma } \approx  0.
\label{thetadot_ls}
\eea
Although they look very similar to the homogeneous equations
(\ref{evolsigma_RW}) and (\ref{thetadot_RW}), these equations are
fully inhomogeneous and encode the evolution of nonlinearities on
large scales. This limit illustrates the separate universe picture
\cite{Sasaki:1995aw,Starobinsky} where the inhomogeneous universe
can be described, on large scales, as juxtaposed Friedmann
homogeneous universes.

If, so far, the order in spatial gradients seems to coincide with
the perturbative classification of the previous subsection, it
differs however for the term
\beq
\nabla^c \sp_c = D^c \sp_c+a^c\sp_c ,
\eeq
which is first order perturbatively but second order in spatial
gradients, at least for the first term on the right hand side,
since $\sp_c$ is already first order in spatial gradients.
 For the second term, it has been shown in \cite{Langlois:2006vv} that $u^a$ can be chosen so that $a^c$ is at least
first order in spatial gradients. We will explicitly verify it in
the next section by working in a coordinate system.

With these prescriptions, the evolution equation of
$\sigma_a^\perp$ and $s_a$ obtained at lowest order in  spatial
gradients become
\bea
&&\left( \ddot \s_a  \right)^{\perp} +\left[\Theta+  \frac{c_s^2}{\PX} \Tdot{\left(\frac{\PX}{c_s^2}\right)} \right]\left( \dot \s_a  \right)^{\perp}
+\frac{c_s^2}{\PX}\left[-P_{,\s\s}-\PX \dot \theta^2+e^{-3 \alpha} \Tdot{\left(  e^{3\alpha} \dot \s P_{,X\s}\right)}    \right]\s_a^{\perp}
+ \dot \s c_s^2 D_a \Theta\nonumber \\
 && \hskip0.3cm \approx \,\hskip0.3cm 
 \Xi \dot s_a +\left[\dot \Xi-\Xi\left(  2\frac{\dot c_s}{c_s} -\frac{P_{,\s}}{\PX \dot \s}       \right)  \right] s_a
  \label{sigma2}
\eea
and
\bea
&& \ddot s_a+\left(\Theta+\frac{\dot \PX}{\PX} \right)\dot s_a+\left(\mu_s^2+\frac{\Xi^2}{c_s^2} \right) s_a
  \approx
-\frac{\Xi}{\PX \dot \s} \epsilon_a
\label{s4}
\eea
where we have dropped the terms containing $\Pi$ in (\ref{s3}), which are two
orders higher than $s_a$ in  spatial gradients.

We now expand the evolution equation for $\zeta_a$,
Eq.~(\ref{zeta_evolution}), by neglecting higher-order spatial
gradients in the two terms on the right hand side of this
equation. The nonadiabatic pressure perturbation becomes
\beq
\Gamma_a  \approx  -\epsilon_a \frac{\dot \s}{\dot \rho}\left[ (1+c_s^2)P_{,\s}-c_s^2 \dot \s^2P_{,X \s} \right]+\PX \dot \s \Xi s_a   , \label{Gamma_ls}
\eeq
while the dissipative nonadiabatic pressure perturbation
$\Sigma_a$ can be completely dropped, since the dissipative term
$\cal D$ is at least second order in the spatial gradients and
thus $\Sigma_a$ is third order in the spatial gradients. Equation
(\ref{zeta_evolution}) therefore becomes, on large scales,
\beq
\dot \zeta_a \approx -\frac{1}{3 \PX^2 \dot \s^3}\left[ (1+c_s^2)P_{,\s}-c_s^2 \dot \s^2P_{,X \s} \right] \epsilon_a
-\frac{ \Theta \Xi}{ 3 \dot \sigma}
 s_a  . \label{zeta_ls}
\eeq
Note that the lowest order limit in  spatial gradients of the
evolution equations of $\sigma_a^\perp$, $s_a$ and $\zeta_a$,
respectively  Eqs.~(\ref{sigma2}--\ref{s4}) and (\ref{zeta_ls}),
are similar to their linear counterparts  except for the terms
$D_a (D^c \sigma_c^\perp)$ and $D_a (D^c s_c)$, which are third
order in  spatial gradients  and therefore negligible in the
spatial gradient expansion. This is because in these equations
the  terms that are higher than linear order in the perturbative
expansion turn out to be  also higher than first order in the
spatial gradient expansion.

We now concentrate our attention on the comoving energy density
perturbation, $\epsilon_a$, defined  in Eq.~(\ref{epsilon}). Until
now we have only made use of the Klein-Gordon equations of the
scalar fields. However, in order to study the behavior of the
comoving energy density, we will now make use of the Einstein
equations, in particular of the so-called {\em constraint
equations}. The projection of the Einstein equations along $u^a$
yields the  {\em energy constraint} in a covariant form,
\beq
u^a G_{ab}u^b=8\pi G \rho.
\eeq
Assuming that the vector field $u^a$ is hypersurface
orthogonal, it is possible to use the Gauss-Codazzi equations
 and
the decomposition
\beq
D_b u_a=\sigma_{ab}+{1\over 3}\Theta h_{ab},
\label{decomposition_spatiale}
\eeq
which is the spatially projected version of (\ref{decomposition})
(with $\omega_{ab}=0$ since
$u^a$ is here hypersurface orthogonal),
in order to  rewrite the energy constraint as
\beq
\frac{1}{2}\left({}^{(3)} \! R+\frac{2}{3}\Theta^2-\sigma_{ab}
\sigma^{ab}\right)=8\pi G\rho, \label{e_c}
\eeq
where ${}^{(3)} \! R$ is the intrinsic  Ricci
scalar of the space-like hypersurfaces orthogonal to $u^a$.

The mixed projection of Einstein's equations yields the covariant
{\em momentum constraint}
\beq
u^b G_{bc}h^c_a= 8\pi G q_a,
\eeq
which can be rewritten, via Gauss-Codazzi relations and
Eq.~(\ref{decomposition_spatiale}), as
\beq
D_b\sigma_a^{\ b}-\frac{1}{3} D_a\Theta=8\pi G q_a. \label{m_c}
\eeq
By combining the energy and momentum constraints, one obtains the
{\it nonlinear covariant} version of the generalized Poisson
equation, which in the linear theory relates the comoving energy
density to the Bardeen's potential defined from the curvature
perturbation. Here one finds
\beq
\frac{1}{2}D_a\left({}^{(3)} \! R -\sigma_{bc}\sigma^{bc}
\right)+\Theta D_b\sigma^b_a=8\pi G \, {\tilde\epsilon}_a,
\label{poisson2}
\eeq
where we have introduced on the right hand side
the quantity
\beq
{\tilde\epsilon}_a \equiv D_a\rho-\Theta
q_a=D_a\rho+\PX \Theta\dot\sigma\sigma_a^\perp. \label{epsilon_ls_tilde}
\eeq
This quantity can be seen as an alternative generalization of the
comoving energy density perturbation, since, in the linear limit,
it is equivalent to $\epsilon_a$ defined in (\ref{epsilon}). In
the fully nonlinear case, the two quantities
in principle are  different: using
(\ref{conservation}) one gets
\beq
{\tilde\epsilon}_a-\epsilon_a=\frac{1}{\dot\sigma}\left(\D-\frac{1}{3}\PX \Theta\Pi\right)\sigma_a^\perp.
\eeq
This difference becomes however negligible on large scales,
\beq
\epsilon_a \approx \tilde \epsilon_a.
\label{eps_eq_eps_tilde}
\eeq

Now, the left hand side of Eq.~(\ref{poisson2}) contains the
projected gradient of the Ricci scalar, $D_a {}^{(3)} \! R$. From
its definition in terms of derivatives of the metric,
it can be shown that, for a perturbed FLRW universe, this term is
of third order in the spatial gradients. Equation (\ref{poisson2})
can thus be used to show that $\epsilon_a$ in
Eqs.~(\ref{zeta_R_two}), (\ref{s4}) and (\ref{zeta_ls}) can be
neglected on large scales, if the shear can also be neglected in
this limit. Indeed, on large scales, the shear rapidly decreases
in an expanding perturbed FLRW universe. Thus, in this limit the
comoving and uniform density integrated expansion
perturbations $\zeta_a$ and $\R_a$ {\em coincide} (up
to a sign),
\beq
\zeta_a + \R_a \approx 0. \label{zeta_R_ls}
\eeq
Furthermore, one can rewrite Eqs.~(\ref{s4}) and (\ref{zeta_ls})
as a closed coupled system of equations, describing the
large-scale nonlinear evolution of adiabatic and entropy
perturbations,

\beq
 \ddot s_a+\left(\Theta+\frac{\dot \PX}{\PX} \right)\dot s_a+\left(\mu_s^2+\frac{\Xi^2}{c_s^2} \right) s_a \approx \, 0
\label{s_evol_ls}
\eeq
and \beq \dot \zeta_a \approx -  \frac{\Theta \Xi}{3 \dot
\s} s_a \label{zeta_evol_ls}.
\eeq

\smallskip

Remarkably, these equations for the covectors
$\zeta_a$ and $s_a$ look very similar to their linear
counterparts, studied in \cite{Langlois:2008mn},
although  they extend them to describe the full nonlinear
evolution for the perturbations.  Despite their compact appearence, they exhibit a quite involved dependence
 on the form of the Lagrangian. For instance, the mass term $\mu_s^2$, 
 which reduces to $V_{ss}- \dot \theta^2$ for standard Lagrangian, also depends in our case on 
 the entropic projection of the mixed derivative $P_{,XI}$, a characteristic shared by the important parameter $\Xi$.

 As explained in Section \ref{sec:general}, we
 stress again that the equations derived in this subsection in the large scale limit are also valid for two-field DBI inflation with the adequate Lagrangian Eq.~(\ref{DBI}).
In the next sections, always following \cite{Langlois:2006vv},
  we will  show
how to translate our results, obtained within the nonlinear
formalism, in the more familiar  coordinate-based approach,
obtaining quite straightforwardly the evolution
equations for the perturbations up to second order.

\section{Linear perturbations}
\label{sec:linear}

In this and next section, 
we  relate the covariant approach to the more familiar
coordinate based formalism. We analyze the linear
perturbations in the present section,  and we then consider
an expansion up to 
second order in the next one.

Let us thus  introduce generic  coordinates $x^\mu=\{t, x^i\}$ to
describe an  almost-FLRW spacetime. Here a  prime will denote a partial
derivative with respect to the cosmic  time $t$, i.e. ${}' \equiv
\partial / \partial t$, since the dot has been used till now to 
denote the Lie derivative
with respect to $u^a$.

The background spacetime is a FLRW
spacetime, endowed with the metric
\beq
ds^2={\bar g}_{\mu\nu}dx^\mu dx^\nu=-dt^2+
a(t)^2\gamma_{ij}dx^i dx^j.
\eeq
At linear order, the  spacetime geometry  is described by the perturbed metric
\beq
ds^2=\left({\bar g}_{\mu\nu}+\delta g_{\mu\nu}\right)dx^\mu dx^\nu,
\eeq
where the components of the metric perturbations can  be written
as
\beq
\delta g_{00}=-2A, \quad \delta g_{0i}=a B_i, \quad \delta
g_{ij}=a^2 H_{ij}.
\eeq
We  decompose, as usual,  $B_i$ and $H_{ij}$ in the forms
\bea
B_i \tackl = \tackr \vec \nabla_i B + B^V_i, \\
\label{H} H_{ij} \tackl = \tackr 
-2\psi \gamma_{ij}+2 \vec\nabla_i\vec\nabla_j E+2
\vec \nabla_{(i}E^V_{j)}+2 { E}^T_{ij},
\eea
where $B_i^V$ and $E^V_i$ are transverse, i.e., 
$\vec \nabla_i B^V{}^i=0=\vec \nabla_iE^V{}^i$, and 
${E}^T_{ij}$ is transverse and traceless, i.e., $\vec
\nabla_i{E}^T{}^{ij}=0$ and $\gamma^{ij}{E}^T_{ij}=0$. Here $\vec\nabla_i$
denotes the three-dimensional covariant derivative with respect 
to the homogeneous
spatial metric $\gamma_{ij}$ (which is also used to lower or raise
the spatial indices). The matter fields are similarly decomposed into a background and a perturbed
part,
\beq
\p^\I(t,x^i)={\bar\p}^I(t)+\delta \p^I(t,x^i).
\eeq

We  now need to specify the components of the unit vector $u^a$, which defines the time
derivation in our covariant approach. At zeroth order, it is
 natural to
take it orthogonal to the homogeneous slices. At first order, \cite{Langlois:2006vv} chose, for simplicity, $u^\mu$ such that $u_i=0$. This implies that, up to first order, the components
of $u^\mu$ are given by
\beq
\label{components_u}
u^\mu=\{1-A, -B^i/a\} ,
\eeq
 and those of the {\it acceleration} vector are given by
\beq
a^\mu=\{0, \vec\nabla^i A /a^2\}\ .
\eeq
This confirms that  $a^\mu$ can be considered
as first order in spatial gradients, in agreement with the assumption of the previous section.

Since the formalism relies on many covectors, it is useful to first consider a generic covector $Y_a$
and work out the  components of its time derivative $\dot Y_a$. To make the explicit calculation, it
is convenient to replace in the definition of the Lie derivative (\ref{Lie})
the covariant derivatives by partial derivatives and  write
\beq
\dot Y_a= u^b \partial_b Y_a + Y_b \partial_a u^b.
\label{Lie2}
\eeq
At zeroth order, the
components of $\dot Y_a$ are simply
\beq
{\bar {\dot Y}}_\mu=\{{\bar Y}_0', 0,0,0\},
\eeq
assuming that the spatial components $\bar{Y}_i$ vanish so as to respect the symmetries of the geometry.
At first order, we get from (\ref{Lie2}) and (\ref{components_u})
\beq
\label{delta_dot_Y}
\delta(\dot Y_0)=\delta Y_0'-(\bar{Y}_0 A)', \qquad
\delta(\dot Y_i)=\delta Y_i'-\bar{Y}_0 \partial_i A.
\eeq

Let us now consider, for the special case of two scalar fields,
the adiabatic and entropic covectors $\sigma_a$ and $s_a$, which
we have introduced earlier.  Note that with the choice of
four-velocity (\ref{components_u}) $\sigma_i^\perp=\sigma_i$. The
background equations of motion can be deduced immediately from
Eqs. (\ref{evolsigma}) and (\ref{thetadot}) and read
\beq
\frac{{\sib}''}{\bc^2} + 3H  {\sib}' +
\frac{\bP_{,X\sigma}}{\bP_{,X}}\,{\sib}'^2-
\frac{\bP_{,\sigma}}{\bP_{,X}}\,=\, 0, \label{evolsigma_b}
\eeq
\beq
\sib'  \thetab' - \frac{\bP_{,s}}{\bP_{,X}}\,=\, 0, \label{thetadot_b}
\eeq
with 
\beq
\sib_0={\bar{\sigma}}'\equiv \sqrt{ \bar{G}_{IJ}   {\bar{\p}^{I'}}   {\bar{\p}^{J'}}   }.
\label{sigma_bar}
\eeq
Note that we used the symbol $\bar \theta'$ to represent the background value of $\dot \theta$. Again, as there is no angle $\theta$, this is merely a notational convenience.\

  From its definition,  Eq.~(\ref{tan_ort1}), one finds that
the {\it spatial} components of $\sigma_a$  at linear order can be
expressed as
\beq
\delta\sigma_i=\bar{e}_{\s I}\partial_i \delta \p^I  
= \partial_i\delta\sigma\,.
\label{dsigma}
\eeq
We use the notation
\beq
\sif\equiv
\bar{e}_{\s I}\partial_i \delta \p^I ,
\label{sif}
\eeq
in agreement with previous studies in the literature.
By using Eq.~(\ref{delta_dot_Y}) with (\ref{sigma_bar}-\ref{dsigma})
one finds that the spatial components of the first and second time derivatives are given by
\beq
\label{sigma_comp}
\delta(\dot\sigma_i)=\partial_i\left({\sif}'-\sib'A\right), \qquad
\delta(\ddot\sigma_i)=\partial_i\left({\sif}''-\sib'A'-2\sib''A\right).
\eeq
The same procedure for $s_a$ gives
\beq
\delta s_i=\partial_i \sf, \qquad
\sf\equiv \bar{e}_{s I}\partial_i \delta \p^I  ,
\eeq
which also coincides with the notation of \cite{Langlois:2008mn}. Since $s_a$, in contrast to $\sigma_a$,
has no longitudinal component, $\sb_0=0$ and the spatial components of $\dot s_a$ and $\ddot s_a$
are simply
\beq
\label{s_comp} \delta(\dot s_i)=\partial_i{\sf}', \qquad
\delta(\ddot s_i)=\partial_i{\sf}''.
\eeq

Plugging the explicit components (\ref{sigma_comp}) and
(\ref{s_comp}) into the linearized equations for $\sigma_a$ and
$s_a$, given by (\ref{sigma1_spatial_lin}) and (\ref{s1_lin})
respectively, one easily obtains the linearized equations for
$\delta\sigma$ and $\delta s$. These read, respectively,
\bea \label{sigma_evol_1}
&& \sif'' +\left[3H+  \frac{\bc^2}{\bP_{,X}}   \left(\frac{\bP_{,X}}{\bc^2}\right)' \right]
 \sif' 
+\frac{\bc^2}{\bP_{,X}}\left[-\bP_{,\s\s}-\bP_{,X} \bar \theta^{'2}  +e^{-3 \bar  \alpha} \left(  e^{3 \bar \alpha}\bar{\sigma}'\bP_{,X\s}\right)'   \right]\sif
\nonumber \\
 && \hskip0.8cm - \bar{\sigma}' \left[
 A'+\left(
 2\frac{\bar{\sigma}''}{\bar{\sigma}'}
+3 H \,\left(1+\bc^2\right)+\frac{\bc^2}{\bP_{,X}}
 \left(\frac{\bP_{,X}}{\bc^2}\right)'
 \right)A+3\bc^2\psi'-\bc^2\vec
\nabla^2(E'-B/a)
 \right]-\frac{\bc^2}{a^2} \vec\nabla^2 \sif \nonumber\\&&=
\bar \Xi \delta s' +\left[ \bar \Xi'-\bar \Xi\left( 2 \frac{\bc'}{\bc} -\frac{\bP_{,\s}}{\bP_{,X} \sib'}      \right)  \right]\delta s
\eea
and
\beq
\sf'' +\left(3H+\frac{\bar{P}'_{,X}}{\bar{P}_{,X}}\right)\sf' +\left( \bar{\mu}_s^2+\frac{\bar \Xi^2}{\bc^2} \right)
\sf-\frac{1}{a^2}\vec\nabla^2\sf =-  \frac{\bar \Xi}{\bar{P}_{,X} \,\bar{\sigma}'
} \delta \epsilon\, ,
\label{s_evol_1}
\eeq
where we have used 
\beq
\bar\Theta=3H, \qquad
\delta\Theta=-3HA-3\psi'+{\vec\nabla}^2(E'-B/a)\,.
\label{delta_Theta}
\eeq

In the  latter equation we have introduced  the first-order
comoving energy density perturbation $\delta\epsilon$, defined by
\beq
\delta \epsilon_i = \partial_i \delta \epsilon, \qquad \delta
\epsilon \equiv \delta \rho - \frac{\rhob'}{\sib'} \sif ,
\eeq
which follows from  the definition (\ref{epsilon}) of $\epsilon_a$.
Using
\begin{eqnarray}
\rhob &=& \bP_{,X} \sib'^2 -\bP\,,\nonumber \\
 \rhof &=&  \frac{\bP_{,X} \,\sib'}{\bc^2}
\left(\sif'-\sib' A- \bar \Xi \sf \right) + \left(\bP_{,X \sigma}\sib'^2-\bP_{,\sigma}\right) \sif,
\end{eqnarray}
one sees that $\delta \epsilon$ can be expressed as
\beq
\delta \epsilon =
\frac{\bP_{,X} \,\sib'}{\bc^2}
\left[  \sif'-\sib' A -\frac{\sib'' }{\sib'} \sif-\bar \Xi \sf \right]\, . 
\label{delta_epsilon_1}
\eeq

Moreover, linearizing the spatial components of the energy
constraint (\ref{e_c}) yields
\beq
3H\left(\psi'+HA\right)-\frac{1}{a^2}{\vec
\nabla}^2\left[\psi+H(a^2E'-aB)\right]=-4\pi G\, \delta\rho,
\label{energy_constraint_1}
\eeq
while the momentum constraint (\ref{m_c}) gives, since $\delta q_i=-
\bP_{,X}\,
\sib'\partial_i\sif$,
\beq
\psi'+HA=4\pi G\,\bP_{,X}\,\sib'\sif.
\label{momentum_constraint_1}
\eeq
Combining the two above constraints yields the
relativistic Poisson-like equation
\beq
\frac{1}{a^2}{\vec \nabla}^2\left[\psi+H(a^2E'-aB)\right]=4\pi G\,
\delta\epsilon, \label{epsilon_ls}
\eeq
which can also be directly obtained by linearizing the spatial
components of Eq.~(\ref{poisson2}). This equation  shows that the comoving
energy density perturbation $\delta\epsilon$ is second order in
the spatial gradients, and thus negligible on large scales in
Eq.~(\ref{s_evol_1}).

 The quantity $\sif$ is not gauge-invariant,
  in contrast with $\sf$.
This is why it is useful to consider the gauge invariant
Sasaki-Mukhanov variable $Q_{\rm SM}$, defined as \cite{SM}
\beq
Q_{\rm SM} \equiv \sif + \frac{\sib' }{H} \psi.
\eeq
Note that the above traditional definition does not follow exactly from the definition
  of  $Q_a$ given earlier in
Eq.~(\ref{Q_two}). Indeed, from $Q_a$,  one can extract a scalar quantity $Q$ defined
as
\beq
Q_i = \partial_i Q, \qquad Q \equiv \sif - \frac{\sib' }{H} \af ,
\eeq
where $\af$ can be written in terms of metric perturbations by
making use of Eqs.~(\ref{alpha_def}) and (\ref{delta_Theta})
(see \cite{Langlois:2005qp}),
\beq
\af = - \psi + \frac{1}{3} \int \vec \nabla^2 (E'-B/a) dt.
\label{alpha_psi}
\eeq
Thus the scalar variable $Q$ coincides with $Q_{\rm SM}$ only in
the large-scale limit.

In the flat gauge\footnote{In the following, a hat indicates that the corresponding quantity is evaluated in the flat gauge.}, defined by $\hat \psi \equiv 0$, $\sif$
coincides with $Q_{\rm SM}$,
\beq
\hat \sif =Q_{\rm SM}.
\eeq
In this gauge, it is possible to use the momentum constraint
equations (\ref{momentum_constraint_1}) to derive the metric
perturbation $A$ as a function of $Q_{\rm SM}$,
\beq
\hat A = -\frac{H'}{H \sib'}Q_{\rm SM}, \label{Af1}
\eeq
and one can write the Poisson equation (\ref{epsilon_ls}) as
\beq
\vec \nabla^2 (\hat E'-\hat B/a) =-\frac{H'}{H\,\bc^2\,\sib'} \left[Q_{\rm
SM}' + \left( \frac{H'}{H} - \frac{\sib''}{\sib'} \right) Q_{\rm SM}
-\bar \Xi \sf \right] , \label{Qf'}
\eeq
where we have used the expression (\ref{delta_epsilon_1})
specialized to the flat gauge. 

By replacing Eqs.~(\ref{Af1}) and
(\ref{Qf'}) into the evolution equation of $\hat \sif$, one
finds the evolution equation of $Q_{\rm SM}$
\cite{Langlois:2008mn},
\bea
Q_{\rm SM}''+ \left(3H+\frac{\bc^2}{\bP_{,X}} \left(\frac{\bP_{,X}}{\bc^2}\right)'\right) Q_{\rm SM}' + \left(\mu_{\s}^2 - \frac{\bc^2 \, \vec \nabla^2}{a^2} \right) Q_{\rm SM} \, = \,  (\bar \Xi \sf)' -\bar \Xi \left( \frac{ \left(H \bc^2\right)'  }{H \bc^2}   - \frac{\bP_{,\s}}{\sib' \bP_{,X}}  \right) \sf\, ,\label{equation_Q_1}
\eea
with
\begin{eqnarray}
\mu_{\s}^2 &\equiv & -\left[\frac{(\sib' /H)'}{ \sib'/H} \right]'-\left(3H+\frac{\bc^2}{\bP_{,X}}\left(\frac{\bP_{,X}}{\bc^2}\right)'+\frac{(\sib'/H)'}{\sib'/H}\right)
\frac{(\sib'/H)'}{\sib'/H}
\label{mu_sigma}\, .
\end{eqnarray}
When one considers only {\em large scales},
the expression (\ref{Qf'}) reduces to
\beq
 Q_{\rm
SM}' + \left( \frac{H'}{H} - \frac{\sib''}{\sib'} \right) Q_{\rm SM}
- \bar \Xi \,
 \sf 
\approx 0,
\label{integral_Q_1}
\eeq
which means that there exists a first integral for the quantity
$Q_{\rm SM}$ and that the second-order equation of motion
(\ref{equation_Q_1}) is not necessary in this limit. In fact, one
can easily check that the large-scale limit of
(\ref{equation_Q_1}) is an automatic consequence of the first
integral (\ref{integral_Q_1}).

Let us now consider  the evolution equation for $\zeta_a$ with two
scalar fields. The  spatial
components of $\zeta_a$, at linear order, are given by
\cite{Langlois:2005qp}
\beq
\delta \zeta_i =\partial_i \zeta, \qquad \zeta \equiv \af
-\frac{H}{\rhob'} \rhof. \label{zeta_first}
\eeq
  From Eq.~(\ref{alpha_psi}) the scalar variable $\zeta$ is thus
related to the Bardeen gauge invariant variable $\zeta_{\rm B}$,
defined as \cite{Bardeen:1980kt,Bardeen:1983qw}
\beq
\zeta_{\rm B} \equiv -\psi - \frac{H}{\rhob'} \rhof,
\eeq
in a similar way to how $Q$ is related to $Q_{\rm SM}$, and on large
scales these two quantities coincide.

According to Eq.~(\ref{delta_dot_Y}), the spatial components of
$\dot\zeta_a$ are
\beq
\delta( \dot \zeta_i) = \partial_i \zeta^{\,\prime}.
\eeq
On large scales, 
one then finds
the relation
\beq
\zeta' \approx -
\frac{H}{\sib'}\,\bar \Xi \, \delta s.\label{linearzetapr}
\eeq
Note also that from Eq.~(\ref{zeta_R_two}) there is
a simple relation between $\zeta$ and $\R$,
\beq
\zeta+\R=-\frac{\ab'}{\rhob'} \delta \epsilon,
\label{relation_R_zeta}
\eeq
which shows that $\zeta$ and $-\R$ coincide on large scales.

In conclusion, all the equations in this section, derived directly
from the covariant formalism, exactly reproduce the linear results
of \cite{Langlois:2008mn}. In particular, according to Eqs.~(\ref{s_evol_1}) and (\ref{equation_Q_1}), whereas entropic perturbations are sensitive to the usual Hubble horizon $(k \approx a H)$, adiabatic perturbations are amplified at \textit{sound} horizon crossing, $c_s k \approx a H$.  As shown in \cite{Langlois:2008wt,Langlois:2008qf}, these equations  are not directly applicable to two-field DBI inflation, whose Lagrangian is not of the form studied in this paper. Despite that, in that situation, it turns out that Eq.~(\ref{equation_Q_1}) for the adiabatic perturbation is still valid and that Eq.~(\ref{s_evol_1}) for the entropy perturbation is minimally modified, simply replacing the gradient term $\frac{1}{a^2}\vec\nabla^2\sf$ by $\frac{\bc^2}{a^2}\vec\nabla^2\sf$, for the adequate Lagrangian Eq.~(\ref{DBI}), with the consequence that both types of perturbations are amplified at \textit{sound} horizon crossing.

 In the next section we turn to the
second-order perturbations,
 and show the power
 of the method we are using to derive new results.

\section{Second order perturbations}
\label{sec:secondorder}
\def\Xf{X_1}
\def\Xs{X_2}
\def\af{{\delta \alpha}}
\def\as{{\delta \alpha}^{(2)}}
\def\sif{{\delta \sigma}}
\def\sis{{\delta \sigma}^{(2)}}
\def\sf{{\delta s}}
\def\ss{{\delta s}^{(2)}}
\def\phf{{\delta \varphi}}
\def\phs{{\delta \varphi}^{(2)}}
\def\phif{{\delta \phi}}
\def\chif{{\delta \chi}}
\def\phis{{\delta \phi}^{(2)}}
\def\chis{{\delta \chi}^{(2)}}
\def\sif{{\delta \sigma}}
\def\phis{{\delta \phi}^{(2)}}
\def\rhof{{\delta \rho}}
\def\rhos{{\delta \rho}^{(2)}}
\def\ps{{\delta P}^{(2)}}
\def\pf{{\delta P}}
\def\ab{\bar{\alpha}}
\def\rhob{\bar{\rho}}
\def\pb{\bar{P}}
\def\sib{\bar{\sigma}}
\def\phib{\bar{\phi}}
\def\chib{\bar{\chi}}
\def\sb{\bar{s}}
\def\thetab{\bar{\theta}}
\def\As{A^{(2)}}
\def\psis{\psi^{(2)}}
\def\Qs{Q^{(2)}}
\def\Qf{Q}
\def\Vi{V_i}

In this section we consider second-order perturbations (for a recent review, see \cite{Malik:2008im}). We will
thus decompose any scalar quantity $Y$ as
\beq
Y(t,x^i) \equiv \bar Y(t)+\delta Y^{(1)}(t,x^i) + \delta
Y^{(2)}(t,x^i) \label{sec_order_dec},
\eeq
where $\bar Y(t)$ is the background part and $\delta Y^{(1)}$ and
$\delta Y^{(2)}$ are respectively the first and second-order
contributions (note that we do not follow here the convention of
including a numerical factor $1/2$ in front of the second-order
contribution). In our subsequent equations, to simplify the
notation, we will often omit the index ${}^{(1)}$ for the
first-order quantities, unless it is required for clarity reasons.

Our main purpose will be to expand our equations governing
$\sigma_a$, $s_a$ and $\zeta_a$ at second order in the
perturbations. We start with a brief discussion on
the gauge invariance of the curvature perturbation.
 As shown in  \cite{Langlois:2005qp}, the second-order expression
of the spatial components $\zeta_i$ can be written, after some
manipulations,  in the form
\beq
\zeta^{(2)}_i=\partial_i\zeta^{(2)}+\frac{\rhof}{\rhob'}
\partial_i\zeta^{(1)}{}'
\label{delta_zeta_2}
\eeq
with
\beq
\zeta^{(2)}\equiv \as -\frac{H}{\rhob'}\rhos
 - \frac{\rhof}{\rhob'} \left[ \zeta^{(1)}{}' +
\frac{1}{2}\left(\frac{H}{\rhob'}\right)'  \rhof \right]
 ,
\eeq
and $\zeta^{(1)}$ being given in Eq.~(\ref{zeta_first}). 
In \cite{Langlois:2006vv} it has been shown
that $\zeta^{(2)}$ is gauge invariant on large scales:
their proof holds identical in our case and we then refer
the reader to this paper for more details. We now
pass to discuss the evolution equations for adiabatic
and entropy fields up to second order, and analyze how
they source the curvature perturbation on large scales.  The
reader
should  keep in mind that the following discussion, restricted to large scales, is applicable to two-field DBI models on using the adequate Lagrangian Eq.~(\ref{DBI}).

\subsection{Adiabatic and entropy fields}

Here we derive the evolution equations for the adiabatic and
entropy field perturbations  at {\em second order}. 
We will restrict ourselves to {\em large scales}, a situation
 in which the equations become more tractable,  
 starting  from the equations expanded in spatial
gradients discussed in Sec.~\ref{sec:ls}. For convenience, we have
collected in the appendix various background and first-order
expressions that will be used in the rest of this section.

Second-order evolution for the perturbations 
 when several scalars are involved has been
first studied
in detail by Malik in \cite{Malik:2005cy} 
  in the large-scale limit
  and with
  canonical
kinetic terms
   (see also
\cite{Enqvist:2004bk,Anupam,Cline}), while
 it has been analyzed
  in \cite{Lyth:2005fi,Vernizzi:2006ve}
 using the separate
universe approach. The second-order decomposition into adiabatic and entropy
components however, has only been derived in  \cite{Langlois:2006vv}, in the case of standard kinetic terms (see \cite{Langlois:2008vk} for a recent application and \cite{Kawasaki:2008sn,Kawasaki:2008pa}
 for related studies). This decomposition
 can easily be derived
 since the fully
nonlinear adiabatic and entropy components
have already been identified. We will
show how to extend this  decomposition into
 adiabatic and entropy components to the case of non-trivial metric in field space, 
and use it to derive the corresponding evolution equations
for these modes. 

In order to pursue the decomposition, 
one starts by expanding the definition of $\sigma_a$ in
Eq.~(\ref{tan_ort1}) at second order. After some straightforward
manipulations, one can write the spatial components of $\sigma_i$
as 
\beq
\delta \sigma_i^{(2)} =\partial_i  \left[   \bar{e}_{\sigma I}  {\delta^{(2)} \phi^I} + \frac{1}{2} \besi^J \bar  \Gamma_{J I K} \left(\besi^K \besi^I \delta \s^2 + \bes^K \bes^I \delta s^2  + 2 \besi^K \bes^I \delta \s \delta s   \right)    \right]
+  \frac{1}{\sib'} \left( \sf'
+  \bar \theta'  \sif \right)\partial_i \sf .  \label{1st_deltasigma}
\eeq
To deal with the term  $\sf' \partial_i \sf$, which cannot be
written as a total gradient, it is convenient to introduce the
spatial vector
\beq
V_i \equiv \frac{1}{2} (\sf \partial_i \sf' - \sf' \partial_i \sf) =
-\sf' \partial_i \sf + \frac{1}{2} \partial_i (\sf \sf'), \label{V_def}
\eeq
which vanishes when $\sf'$ and $\sf$ have the same spatial
dependence, i.e., $\sf'=f(t)\sf$.

 By expanding also
the definition of $s_a$ in  Eq.~(\ref{tan_ort2}), one finds,
for $s_i$ and $\sigma_i$, respectively, 
\bea
\label{sigi2} \delta \sigma_i^{(2)} \tackl = \tackr \partial_i
\sis + \frac{\thetab'}{\sib'} \sif\partial_i \sf -\frac{1}{\sib'} V_i, \label{si_i} \\
\label{s_i}  \delta s_i^{(2)}  \tackl = \tackr \partial_i \ss +
\frac{\sif}{\sib'}  \partial_i \sf',
\eea
with
\bea
\label{si2} \sis \tackl \equiv \tackr   \bar{e}_{\sigma I}  {\delta^{(2)} \phi^I} + \frac{1}{2} \besi^J \bar \Gamma_{J I K} \left(\besi^K \besi^I \delta \s^2 + \bes^K \bes^I \delta s^2  + 2 \besi^K \bes^I \delta \s \delta s   \right)   + \frac{1}{2 \sib'} \sf \sf',  \\
\ss \tackl \equiv \tackr   \bar{e}_{s I}  {\delta^{(2)} \phi^I} + \frac{1}{2} \bes^J \bar  \Gamma_{J I K} \left(\besi^K \besi^I \delta \s^2 + \bes^K \bes^I \delta s^2  + 2 \besi^K \bes^I \delta \s \delta s   \right)   -\frac{\sif}{\sib'}  \left( \sf' +
\frac{\bar \theta'}{2}
 \sif\right).\label{sis}
\eea
The form of the right hand side of Eq.~(\ref{s_i}) has been chosen
by analogy with the form (\ref{delta_zeta_2}). Since $s_a$
vanishes at zeroth order, arguments similar to those used
in \cite{Langlois:2006vv} to prove gauge invariance
for $\zeta^{(2)}$  ensure that $\ss$, defined in Eq.~(\ref{sis}),
is also gauge invariant on large scales. The form of $\ss$ is,
then, forced by our  covariant definition. Let us remark that the 
non-flat nature of the field space metric manifests itsef in
the appearance of the terms with Christoffel symbols in 
$ \sis$ and $\ss$. Otherwise, these are formally identical to the corresponding 
definitions in \cite{Langlois:2006vv}.

Note that $\delta s_i^{(2)}$ contains the first-order adiabatic
perturbation. This is due to the fact that the adiabatic and
entropy components are defined locally: whereas the first-order
components are defined with respect to a background basis in field
space, which is {\em only} time dependent, the second-order
components will be sensitive to the first-order fluctuations of
the field space basis, which can be expressed in terms of the
first-order adiabatic and entropy components. 
Then, 
the adiabatic
component $\sigma_a$ does not vanish at zeroth order and $\sis$ is
not a gauge invariant variable. It
is then  useful to consider our generalization of the Sasaki-Mukhanov
variable, $Q_a$,  defined in Eq.~(\ref{Q_two}). Its spatial
components can be expanded at second order in the perturbations,
similarly to what we have done with $\zeta_a$,
\beq
Q_i^{(2)} = \partial_i \Qs + \frac{\af}{H}  \partial_i \Qf'
+  \frac{\thetab'}{\sib'} \Qf\partial_i \sf -\frac{1}{\sib'}V_i, \label{Q2_i}
\eeq
where $\Qs$ is defined as
\beq
\Qs \equiv \sis - {\sib'\over H}\as
 - \frac{\af}{H} \left[ Q^{(1)}{}' + \frac{1}{2}
 {\left({\sib'\over H}\right)}' \af - \thetab' \sf \right]
\label{Q_2}.
\eeq
From this expression  it is natural to define
\beq
Q_{\rm SM}^{(2)} \equiv \sis + {\sib'\over H} (\psis+ \psi^2)
 + \frac{\psi}{H} \left[ Q_{\rm SM}^{(1)}{}' - \frac{1}{2}
 {\left({\sib'\over H}\right)}'
 \psi -  \thetab' \sf \right]
\label{QSM_2}
\eeq
as the  local part of the scalar gauge invariant second-order
Sasaki-Mukhanov variable.  Restricted to a single scalar field,
this definition coincides with the one given in
\cite{Malik:2005cy}.
 Note that we cannot write Eq.~(\ref{Q2_i})
in the same form as Eq.~(\ref{delta_zeta_2}) because the last two
terms on the right hand side cannot be written as a total spatial
gradient.

The second-order (in time) evolution of $\sigma_a$ is given by
Eq.~(\ref{sigma2}). However, on large scales we do not need to
compute a second-order differential equation because the
adiabatic evolution is governed by a first integral, as in the
linear case. This first integral is obtained directly from the
constraint equations and it is not necessary to expand
(\ref{sigma2}) at second order in the perturbations.

In order to compute this first integral we need  the
second-order energy and momentum constraints, which can be derived
by expanding Eqs.~(\ref{e_c}) and (\ref{m_c}) and by using 
\beq
\delta \Theta^{(2)} \approx  \frac{9}{2} H A^2 +3A \psi' -6 \psi
\psi'-3HA^{(2)}-3\psi^{(2)'}.
\label{delta_theta_2}
\eeq
On large scales, one can write the second-order energy constraint
equation as {\cite{Langlois:2006vv}}
\beq
3 H\left[H \As +  \psi^{(2)}{}' +2  \psi \psi'
- \frac{1}{2} \frac{\psi'{}^2}{H}
-2 A(HA+ \psi' )   \right] \approx -4
\pi G \rhos,
\label{energy_2}
\eeq
where $\rhos$ can be 
 calculated using (note that the 
non-flat nature of the field space metric simply manifests itsef in
the presence of the Riemann tensor)
 
 \begin{eqnarray}
 \delta^{(2)} \dot \sigma^2
&=&
{\overline \sigma'}^{2}\left(4A^2-2 A^{(2)} \right)-4A \overline \sigma' \left( \delta \s'-\overline \theta' \delta s \right)+ \left( \delta \s'-\overline \theta' \delta s \right)^2+  \left( \delta s' + \overline \theta' \delta \sigma \right)^2 
 \nonumber \\
 &-& 2  \overline \sigma'   \overline \theta' \delta s^{(2)} +2  \overline \sigma'  \sis{}' -2  \overline \theta' \delta \s \left( \delta s' +\frac{ \overline \theta' }{2}\delta \sigma  \right)- \overline \sigma'  \left(\frac{1}{ \overline \sigma' } \delta s \delta s' \right)'
  \nonumber \\
 &+& {\overline \sigma'}^{2} \besi^I \besi^J \bes^K \bes^L \bar R_{K I J L}\delta s^2\,.
 \label{ds2}
  \end{eqnarray} 
  
Explicitely, this gives
\bea
\rhos \tackl \approx \tackr \frac{\bP_{,X}}{\bc^2}  \left( \sib' \sis{}' - {\overline \sigma'}^{2} \As - \sib'  \bar \Xi \delta s^{(2)}  \right)
+\left({\overline \sigma'}^{2} \bar{P}_{,X \s}-\bar P_{,\s} \right) \sis + \Lambda_\rho
\eea
where $\Lambda_\rho$ is a quadratic function of $A$,  $\delta
\sigma^{(1)}$, $\delta s^{(1)}$, and their first derivatives,
\bea
\Lambda_\rho\tackl  \equiv  \tackr \frac{\bar P_{,X}}{2 \bc^2} \sif' (\sif'-2
\thetab' \sf) + \frac{ \sf'}{\sib'} \left[\frac{1}{2  \bc^2}\left((1+\bc^2)\bar{P}_{,\s}-\bc^2 {\overline \sigma'}^{2} \bar P_{,X \s} \right)
\sf - \left(\bar{P}_{,s}-{\overline \sigma'}^{2} \bar P_{,Xs}   \right)
\sif\right]
\nonumber \\
\tackl -\tackr  \frac{1}{2}\left[\bar P_{,\sigma \sigma} - {\overline \sigma'}^{2} \bar P_{,X\s \s}
+\frac{\thetab'}{\sib'}\left(\bar P_{,s} - {\overline \sigma'}^{2} \bar P_{,X s} \right) \right] \sif^2  +\left({\overline \sigma'}^{2} \bar P_{,X \s s}-\bar P_{, \s s}   \right)\sf \sif
\nonumber \\
\tackl +\tackr
\frac{1}{2}\left[ {\overline \sigma'}^{2} \bar P_{,X s s}-\bar P_{, s s}+\frac{\bar P_{,X}}{\bar c_s^2}\left(\bar  \mu_s^2+\frac{\bar{\Xi}^2} {\bc^2} + {\overline \theta'}^{2}+{\overline \sigma'}^{2} \besi^I \besi^J \bes^K \bes^L \bar R_{K I J L} \right)  \right] \delta s^2
\nonumber \\
\tackl +\tackr
\frac{1}{2} \left(\bar P_{,XXX} {\overline \sigma'}^{4} +3 \bar P_{,XX} {\overline \sigma'}^{2} \ \right) \left( \delta \s'- \overline \theta' \sf - A \overline \sigma' \right)^2
\nonumber \\
\tackl +\tackr
\overline \sigma'  \left[ \left( \bar P_{,XX\s} {\overline \sigma'}^{2} +\bar P_{,X \s} \right) \delta \s + \left( \bar P_{,XXs} {\overline \sigma'}^{2} +\bar P_{,X s} \right) \delta s -\frac{2 \bar P_{,X}}{\bar c_s^2} A \right] \left( \delta \s'- \overline \theta'  \sf - A \overline \sigma' \right).
\eea
The second-order momentum constraint equation reads {\cite{Langlois:2006vv}}
\beq
\partial_i \left[H \As + \psi^{(2)}{}' + 2 \psi \psi' - \frac{1}{2}HA^2
-A(HA+\psi') \right]
\approx -4 \pi
G \delta q^{(2)}_i,
\label{momentum_2}
\eeq
where the second-order momentum $\delta q^{(2)}_i$ is given, from
Eq.~(\ref{qa}), by
\beq
\delta q_i^{(2)} =-\partial_i \left[\bP_{,X} \sib' \sis
+\frac{\bP_{,X \sigma} \sib' }{2} \delta \sigma^2
+\frac{\bP_{,X}}{2 \,\bc^2} \frac{\sib''}{\sib'} \sif^2 + \bP_{,X} \thetab' \sif \sf
\right] -\frac{1}{\sib'}
\delta \epsilon \partial_i \sif +\bP_{,X}\,V_i.
\label{momentum_2_def}
\eeq
As already noticed in \cite{Rigopoulos:2005xx} for large scales,
 $\delta q_i^{(2)}$ {\em cannot} be written as a total gradient when several
scalar fields are present. After neglecting $\delta \epsilon$ on
large scales in  the above equation, this is manifest because of
the presence of $V_i$. This implies that,  in principle, if $V_i$
does not vanish,  one cannot define at second-order a comoving
gauge,  i.e., such that $\delta q_i^{(1)}=0$ {\em and} $\delta
q_i^{(2)}=0$, in  contrast with the linear theory or the
single-field case.

However, it is instructive to  derive the  evolution equation for $V_i$ on
large scales by using the linear evolution equation for $\sf$,
Eq.~(\ref{s_evol_1}), neglecting the gradient term and $\delta
\epsilon$ at first order. One finds a simple expression that generalizes the corresponding result of \cite{Langlois:2006vv} to the case
of non-canonical kinetic terms
\beq
V_i'+ \left(3 H +\frac{\bP_{,X}'}{\bP_{,X}}\right) V_i=0.
\label{evolution_Vi}
\eeq
This implies that, in an expanding universe, $\bP_{,X} V_i$ will  decay
like $a^{-3}$
 and rapidly become negligible even if it is nonzero
initially.  Consequently, in
an expanding universe one can in practice  ignore $V_i$ on large
scales and thus define,  {\em in an approximate sense}, a comoving
gauge at second order, which coincides with $\sif^{(1)}=0=\sis$.
In this approximate comoving gauge, the momentum
(\ref{momentum_2_def}) can be written as a total gradient. In the
rest of the paper, in order to remain as general as possible, we 
will keep the term $V_i$.

Similarly to the first-order case, it is possible to combine the
energy and momentum constraint equations and derive the
relativistic Poisson-like equation analogous to
Eq.~(\ref{epsilon_ls}), which corresponds to the expansion, at
second order and on large scales, of Eq.~(\ref{poisson2}). By
expanding Eq.~(\ref{epsilon_ls_tilde}) and using
(\ref{eps_eq_eps_tilde}), one has
\beq
\delta \epsilon^{(2)}_i \approx \partial_i \rhos - 3H \delta q_i^{(2)}
- \delta \Theta \delta q_i \approx 0, \label{Poisson_2}
\eeq
where the last approximate equality is a consequence of
Eqs.~(\ref{delta_Theta}), (\ref{momentum_constraint_1}), (\ref{energy_2}) and (\ref{momentum_2}) and confirms our
conclusion of  Sec.~\ref{sec:ls} in a covariant context, namely
that we can neglect $\epsilon_a$ on large scales. The second-order
spatial components of  $\epsilon_a$ defined in
Eq.~(\ref{epsilon}), can be decomposed as
\beq
\delta \epsilon_i^{(2)} = \partial_i \delta \epsilon^{(2)} +
\frac{\sif}{\sib'}
\partial_i \delta \epsilon^{(1)}{}' -3H\,\bP_{,X}\, V_i, \label{epsilon_i2}
\eeq
with $\delta \epsilon^{(2)}$ defined by
\beq
\delta \epsilon^{(2)} \equiv \rhos - {\rhob'\over \sib'} \sis
 - \frac{\sif}{\sib'} \left[ \delta \epsilon^{(1)}{}'  +\frac{1}{2}
 {\left({\rhob'\over \sib'}\right)}'
 \sif  +    \frac{\rhob'}{\sib'} \bar \theta' \sf \right]
 . \label{delta_epsilon_def_2}
\eeq
It is only when $V_i$ is negligible that  the quantity $\delta
\epsilon^{(2)}$ can be interpreted as  the comoving energy density
at second order. Otherwise, as discussed before, the comoving
gauge cannot be defined.

Using  the decomposition (\ref{epsilon_i2}) and the fact that
$\delta \epsilon^{(1)}$ is negligible on large scales,
Eq.~(\ref{Poisson_2}) can be written as
\beq
\partial^{2} \delta \epsilon^{(2)} \approx 3H \bP_{,X}
\partial^iV_i.
\label{epsilon2_ls}
\eeq
When $V_i$ is negligible, and only then, one finds that, like at
first order, the second-order comoving energy density is
negligible on large scales,
\beq
\delta \epsilon^{(2)} \approx 0.
\eeq

Having discussed the general properties of the second-order
constraint equations, we can now derive the evolution equation of
the gauge-invariant adiabatic component $Q_{\rm SM}^{(2)}$.
Similarly to what we have shown in the previous section for the
first-order variables, the simplest way to derive  an equation
satisfied by $Q_{\rm SM}^{(2)}$ is to work in the flat gauge $\hat
\psi^{(1)}= 0 =\hat \psi^{(2)}$, where $\sis$ reduces to $Q_{\rm
SM}^{(2)}$,
\beq
\label{Q_3} \hat \sis = Q_{\rm SM}^{(2)}.
\eeq
In this gauge, Eqs.~(\ref{energy_2}) and (\ref{momentum_2})
reduce, respectively, to
\bea
3 H^2\left(\hat  \As  -2 \hat  A^2  \right) \tackl \approx \tackr-4
\pi G \hat \rhos, \\
H \partial_i \left( \hat  \As  - \frac{3}{2} \hat  A^2 \right)
\tackl \approx \tackr -4 \pi
G \hat{\delta q}^{(2)}_i.
\eea 
Using the first-order constraint equations Eqs.~(\ref{Af1}) and
(\ref{integral_Q_1}), these can be rewritten as
\bea
\hat \As \tackl \approx \tackr 
\frac{1}{3 H^2+H'/\bar c_s^2} \left[-\frac{\sib' \bar P_{,X}}{2 \bar c_s^2}\left( Q^{(2)}_{\rm SM}{}'-\bar \Xi \delta s^{(2)} \right)
+ \frac 1 2 \left(\bar P_{,\s}-\sib'{}^{2} \bar P_{,X \s} \right)  Q^{(2)}_{\rm SM} \right. +
\left. 6 \left( \frac{H'}{\sib'} \right)^2  Q_{\rm SM}^2
-\frac{\hat \Lambda_\rho}{2} \right]
\label{c_1}
\eea
and
\beq
\hat \As \approx -\frac{H'}{H \sib'}  \left[ \Qs_{\rm SM} +
\frac{\thetab'}{\sib'} Q_{\rm SM} \sf + \frac{1}{2 \sib'} \left(
\frac{\sib''}{ \bar c_s^2 \sib'} -3 \frac{H'}{H}+\frac{\sib' \bar P_{,X \s}}{\bar P_{,X}} \right) Q_{\rm SM}^2 -
\frac{1}{\sib'} \partial^{-2}
\partial^i V_i\right]  . \label{c_2}
\eeq
 This last equation
contains a nonlocal term because we have written the momentum
constraint as a scalar equation while keeping the second-order
vector $V_i$ defined in (\ref{V_def}). By combining these two
relations to get rid of $\As$ one obtains the following first
integral for $Q^{(2)}_{\rm SM}{}$:
\bea
Q_{\rm SM}^{(2)}{}' \tackl + \tackr
\left(\frac{H'}{H}-\frac{\sib''}{\sib'} \right) Q_{\rm SM}^{(2)}
-\bar \Xi  \ss \nonumber \\ \tackl \approx \tackr   \left[
3 \sib' \bar P_{,X} \bar c_s^2 - \frac{1}{2 \sib'}
\left(\frac{\sib''}{\bar c_s^2 \sib'} -3 \frac{H'}{H} +\frac{\sib' \bar P_{,X \s}}{\bar P_{,X}} \right)\left(3 H \bar c_s^2 +\frac{H'}{H} \right) \right] Q_{\rm SM}^2
- \frac{ \bar c_s^2 \hat \Lambda_\rho}{\bar P_{,X} \sib'}
\nonumber \\  \tackl - \tackr \left( 3 H \bar c_s^2 +\frac{H'}{H}  \right)\left( \frac{\thetab'}{\sib'} Q_{\rm SM} \sf -
\frac{1}{\sib'} \partial^{-2} \partial^i V_i  \right).
\label{integral_Q_2}
\eea
This equation is the second-order equivalent of
(\ref{integral_Q_1}).  We have put together on the left hand side
all the terms depending on purely second-order quantities, and on
the right hand side all the terms which are quadratic in
first-order quantities. As in the first-order case, the entropy
perturbation sources the evolution of the adiabatic perturbation.
The nonlocal term containing $V_i$ comes from the momentum
constraint and is a new feature with respect to the first-order
case (or the second-order case for a {\em single} scalar field).
However, as we have discussed earlier, it becomes quickly
negligible on large scales in an expanding universe, in which case
the first integral (\ref{integral_Q_2}) becomes a scalar local
equation.

\smallskip

We eventually derive a second-order equation
for the entropy mode $\ss$, which will be second order in time. Since we are
restricting ourselves to {\em large scales}, we simply expand the
spatial components of Eq.~(\ref{s4}) up to second order. This
gives 
\bea
&&
\delta (\ddot s_i)^{(2)}+\left(3 H+\frac{ \bar{P}_{,X}^{'}} {\bar{P}_{,X}} \right)
 \delta (\dot s_i)^{(2)}+\left(\bar{\mu}_s^2+\frac{\bar \Xi^2}{\bar c_s^2} \right)
 \delta s_i^{(2)} +\delta \left(
\Theta +
\frac{\dot \PX}{\PX} \right)
\partial_i \sf' 
\cr 
&& \quad 
+ 
\delta \left( \mu_s^2+  \frac{\Xi^2}{c_s^2}\right) \partial_i \sf
\approx -\frac{\bar \Xi}{\bP_{,X} \sib'} \delta \epsilon^{(2)}_i ,
\label{pre_s_evol_2}
\eea
where we have neglected the gradient of the comoving energy
density at first order, $\delta \epsilon_i^{(1)}$ which according
to Eq.~(\ref{epsilon_ls}) is subdominant on large scales.

To proceed, we need the spatial components of the first and second
time derivatives of the covectors $s_a$. By using Eq.~(\ref{Lie2})
for $s_i$ at second order
and ignoring the higher-order terms in
the gradient expansion,  one obtains
\beq
\label{dotsi2} \delta (\dot s_i)^{(2)} \approx \delta
s_i^{(2)}{}'- A
\partial_i \sf',
\eeq
and, by applying once more (\ref{Lie2}),
\beq
\label{ddotsi2} \delta (\ddot s_i)^{(2)} \approx  \delta
s_i^{(2)}{}'' - A'
\partial_i \sf' -2 A
\partial_i \sf'' .
\eeq
Recalling that 
$$\delta s_i^{(2)}   =  \partial_i \ss +
\frac{\sif}{\sib'}  \partial_i \sf'\,,$$
we can then substitute the expressions (\ref{s_i}) and
 (\ref{dotsi2}--\ref{ddotsi2}) for the second-order entropy
component and its derivatives, and 
 use equation (\ref{s_evol_1}) and its
time derivative. Defining 
\beq
\M_{IJ}\, \equiv \, -\frac{ \mathcal D_I \mathcal D_J P}{P_{,X}}+\dot \s^2 R_{IKJL} e_{\sigma}^K e_{\sigma}^L
+\frac{1}{\dot \s^2 P_{X}^2}
\left[
(2+c_s^2) P_{,I} P_{,J}+ c_s^2 \dot \s^4 \,P_{,X I} P_{,X J}
- c_s^2 \dot \s^2 P_{,X I} P_J-c_s^2 \dot \s^2 P_{,X J} P_I
\right]
\eeq
such that
\beq
\mu_s^2+\frac{\Xi^2}{c_s^2}=e_s^I e_s^J \M_{IJ}\,,
\eeq
the equation (\ref{pre_s_evol_2}) can be written as the spatial gradient of the following scalar equation
\bea
&&
\delta s^{(2)''} +\left(3 H+\frac{\bar{P}_{,X}'}{\bar{P}_{,X}} \right) \delta s^{(2)'} +\left(\bar  \mu_s^2+\frac{\bar{\Xi}^2} {\bc^2} \right) \delta s^{(2)}  \approx - \frac{\overline \theta'}{\overline \sigma'}\delta s'^2\, -\left( 3 H\frac{ \bar \Xi}{2\overline \sigma'} -\frac{2}{\overline \sigma'} \besi^I \bes^J \bar \M_{IJ} \right) \delta s  \delta s'\
\cr &&  -
\frac{\bes^I \bes^J}{2} \left[
\overline \sigma'(\bar \Xi-\bar{\theta'})\, \bar \M_{IJ,X}
+ \bes^K\, \mathcal{D}_K \bar \M_{IJ}- 2 \, \frac{\bar \Xi}{\overline \sigma'}\, \bar \M_{IJ}
\right] \delta s^2 \, -\frac{\bar \Xi}{\bar P_{,X} \overline \sigma' } \delta \epsilon^{(2)}\,.
\eea
Let us make a few clarifying comments: when calculating the various derivatives of $\M_{IJ}$ with respect to the fields and to $X$, $\dot \s^2$ is replaced by $2X$ and the basis vectors $\besi^I, \bes^I$ are considered as constant (they are not \textit{functions} of $X$ and the fields); eventually, $ \mathcal{D}_K \bar \M_{IJ} \equiv \partial_K \bar \M_{IJ} -\bar \Gamma_{K I}^L \bar \M_{LJ}-\bar \Gamma_{K J}^L \bar \M_{I L}$.

\smallskip

Note that this equation is closed, in the sense that only the
entropy field perturbation appears: even when $\delta
\epsilon^{(2)}$ is not negligible, it  can be written in terms 
of $\sf$ and $\sf'$ by using Eqs.~(\ref{epsilon2_ls}) and
(\ref{V_def}). Thus, on large scales the entropy field evolves
independently of the adiabatic components, as in the linear
theory.

\bigskip

\subsection{Generalized uniform density and comoving curvature perturbations}\label{last}

We now derive the large-scale evolution equation for
$\zeta^{(2)}$, expanding to second order Eq.~(\ref{zeta_ls}). In
the fluid description, it was shown in
\cite{Malik:2003mv,Langlois:2005qp} that on large scales the
evolution equation for $\zeta^{(2)}$ can be written as
\beq
\zeta^{(2)}{}' \approx - \frac{H}{\bar \rho+ \bar P} \Gamma^{(2)}
- \frac{1}{\bar \rho+ \bar P} \Gamma_1 \zeta_1'   ,
\label{conserv_zeta2}
\eeq
where $\Gamma^{(2)}$ can be read from the second-order
decomposition of the quantity $\Gamma_a$ for a fluid, defined in
Eq.~(\ref{Gamma_def}), i.e.,
\beq
\delta \Gamma_i^{(2)} =
\partial_i  \Gamma^{(2)} + \frac{\rhof}{\rhob'}
\partial_i \Gamma^{(1)}{}'. \label{so_form}
\eeq
In the two-scalar field case considered here we must compare this
expansion with the expression for $\Gamma_a$ given in the
large-scale limit by Eq.~(\ref{Gamma_ls}). Expanding this equation
to second order, using Eq.~(\ref{s_i}), one
obtains
\begin{eqnarray}
\delta \Gamma_i^{(2)} &\approx& 
-\delta \epsilon_{i}^{(2)}
\left\{ \frac{\overline \sigma'}{\overline \rho'}
\left[ \left(1+\bc^2\right)\bP_{,\sigma}-\bc^2 \overline\sigma'^2
\bP_{,X \sigma}
\right]
 \right\}+\left( \bP_{,X} \overline{\sigma}' \,\bar \Xi \right)
 \left( 
 \partial_i \delta s^{(2)} +\frac{\delta \sigma}{\overline{\sigma}'}
 \partial_i \delta s
 \right) \nonumber\\
 &+&  \partial_i \delta s \,\delta\left( P_{,X} \dot{\sigma} \,\Xi\right)
\end{eqnarray}
 where we have used that $\delta \epsilon^{(1)}$ is negligible
on large scales. From this, defining $\X_I$ as
\beq
\X_I \, \equiv \, (1+c_s^2)P_{,I}-c_s^2 \dot{\sigma}^2 P_{,XI}
\eeq
such that
\be
 P_{,X} \dot{\sigma} \,\Xi
  \,=\,e_s^{I} \,\X_I\,,
  \ee
one obtains the expression
\begin{eqnarray}
\Gamma^{(2)} \,\approx\,- \besi^I \bar \X_I\, \left(\frac{\overline \sigma'}{\overline\rho'}
\delta \epsilon^{(2)}+\frac{1}{2 \overline\sigma'}\,
\delta s \delta s'
\right)\, + \bes^I \bar \X_I\,
 \delta s^{(2)} +\left[
\bes^I \bes^J \bar \X_{I;J}
  -\overline \sigma'(\overline \theta'-\bar \Xi) \bes^I \bar \X_{I,X}
 \right]\, \frac{\delta s^2}{2} \,.
\end{eqnarray}
 From the previous expression it is immediate
to  rewrite the second-order evolution equation for
$\zeta^{(2)}$ on large scales,
 Eq.~(\ref{conserv_zeta2}),
 as 
\beq
\zeta^{(2)}{}'\approx -\frac{H}{\bP_{,X}\, \sib'^2} \left[- \besi^I \bar \X_I\, \left(\frac{\overline \sigma'}{\overline\rho'}
\delta \epsilon^{(2)}+\frac{1}{2 \overline\sigma'}\,
\delta s \delta s'
\right)\, + \bes^I \bar \X_I\,
 \delta s^{(2)} +\left(
\bes^I \bes^J \bar \X_{I;J}
  -\overline \sigma'(\overline \theta'-\bar \Xi) \bes^I \bar \X_{I,X}-2 \bar{P}_{,X} \,\bar \Xi^2
 \right)\, \frac{\delta s^2}{2}\right].
 \label{zeta2_evol}
\eeq

The previous expression shows that the second-order curvature perturbation is, as expected,
 sourced only
by entropy modes on large scales.  On the other hand, it
depends in  a much richer way on $\X_I$ and on the
derivatives of $P$, that is on the 
 form of the kinetic terms, than its linear 
 counterpart in eq. (\ref{linearzetapr}). Moreover, $\X_I$ reduces to $-2 V_{,I}$ for a standard Lagrangian, with no dependence on the kinetic term $X$. 
The appearence of $\bar{\X}_{,IX}$ on the right hand side of (\ref{zeta2_evol}), acting as a source for the large-scale evolution of the curvature perturbation, is thus solely due to the non-standard nature of the Lagrangian we are considering.
It is important
to emphasize that the nonlinear
formalism we adopted, as well as the fact that we considered a very general Lagrangian, have allowed us to straightforwardly
 obtain a relatively
elegant and compact expression. The same would be
difficult to get working directly in a coordinate based approach, or considering a specific Lagrangian.

\smallskip

It is also useful to express our results in terms of $\R_a$. The
spatial components of $\R_a$ can be decomposed as
\beq
\R_i^{(2)} = \partial_i \R^{(2)} + \frac{ \sif}{\sib'}
\partial_i \R^{(1)}{}' -  \frac{H}{\sib'{}^2}V_i , \label{R_2_i}
\eeq
with
\beq
\R^{(2)} \equiv-  \as + {H\over \sib'}\sis
 + \frac{\sif}{\sib'} \left[ -\R^{(1)}{}' + \frac{1}{2}
 {\left({H\over \sib'}\right)}'
 \sif  +  \bar \theta' \frac{H}{\sib'}
 \sf \right]
\label{R_2}.
\eeq
The last term in Eq.~(\ref{R_2_i}) comes from the fact that, like
$\epsilon_a$ and in contrast to $\zeta_a$, $\R_a$ is defined in
terms of the spatial momentum which cannot be expressed in general
as a pure gradient. When this term can be neglected, and only
then, $\R^{(2)}$ coincides with the second-order comoving
curvature perturbation defined in
\cite{Maldacena:2002vr,Vernizzi:2004nc}.

It is easy to derive a first-order (in time) evolution equation
for $\R^{(2)}$ by noting that $\zeta^{(2)}$ and $\R^{(2)}$ are
related on large scales. Indeed, expanding Eq.~(\ref{zeta_R_ls})
to second order using (\ref{delta_zeta_2}) and (\ref{R_2_i}), and
neglecting terms proportional to $\delta \epsilon^{(1)}$, one gets 
\beq
\zeta^{(2)} +\R^{(2)} \approx  \frac{1}{3 \bP_{,X} \sib'{}^2} \delta
\epsilon^{(2)}. \label{R_zeta_rel2}
\eeq
When $\delta \epsilon^{(2)}$ is negligible on large scales, like
in an expanding universe where we can neglect $V_i$, $\zeta^{(2)}$
and $\R^{(2)}$ coincide on large scales as in the single-field
case \cite{Vernizzi:2004nc}. However, this is not true in general
in the multi-field case if $V_i$ cannot be neglected.

\bigskip

From this relation and the evolution equation of $\zeta^{(2)}$,
Eq.~(\ref{zeta2_evol}), one can find a large-scale evolution
equation for $\R^{(2)}$,

\beq
\R^{(2)}{}' \,\approx \, \frac{H}{ \bP_{,X} \sib'^2} \left[ - \frac{\besi^I \bar \X_I}{2 \overline\sigma'}\, \delta s \delta s'  + \bes^I \bar \X_I\,
 \delta s^{(2)} +\left(
\bes^I \bes^J \bar \X_{I;J}
  -\overline \sigma'(\overline \theta'-\bar \Xi) \bes^I \bar \X_{I,X}-2 \bar{P}_{,X} \,\bar \Xi^2
 \right)\, \frac{\delta s^2}{2} +(\bc^2+\frac{H'}{3  H^2})\delta \epsilon^{(2)} \right]. 
\label{R2_evol}
\eeq
 The second-order uniform adiabatic
field perturbation $\R^{(2)}$ can be related on large scales to
$Q_{\rm SM}^{(2)}$, by combining Eqs.~(\ref{QSM_2}) and
 (\ref{R_2}).
 One obtains
\beq
\R^{(2)} \approx \frac{H}{\sib' } \left[Q_{\rm SM}^{(2)}
-\frac{1}{\sib'} \left(Q_{\rm SM}'- \thetab'\sf \right) Q_{\rm SM}
- \frac{1}{2H} \left(\frac{H}{\sib'} \right)' Q_{\rm SM}^2
\right], \label{RQ_rel}
\eeq
 which can be
used, together with the linear first integral
(\ref{integral_Q_1}), to show that Eq.~(\ref{R2_evol}) is
equivalent to the first integral (\ref{integral_Q_2}).

\section{Conclusions}
\label{sec:conclusion}

In this work, we adopted a covariant formalism to 
derive exact evolution equations for nonlinear
perturbations, in a universe dominated by two scalar
fields. These scalar fields are characterized
by  non-canonical kinetic terms 
 and  an 
arbitrary field space metric, 
a situation typically encountered in inflationary
models inspired by string theory. Our  exact equations can be expressed in a 
relatively compact way, and in physically interesting limits they
closely resemble their linear counterpart.
 They acquire a quite simple and elegant form, due to the fact that 
 we considered a very general Lagrangian to perform our calculations.
 
\smallskip

Extending the methods of \cite{Langlois:2006vv} to a general field space metric,
we have decomposed our non-linear scalar perturbations
into adiabatic and entropy modes, corresponding to a generalization 
of analogous definitions adopted in the linear theory. Then, we
derived the corresponding evolution equations, that acquire several
new contributions associated with the non-canonical kinetic terms
for the scalar fields. We also obtained a nonlinear generalization
of the curvature perturbation on uniform energy density hypersurfaces, 
 showing that
on large scales it is sourced only by the nonlinear entropy perturbation.

\smallskip

We have used these nonlinear equations as a starting point to show
how to extend and generalize some
  results previously obtained in the literature, in
   a relatively 
  straightforward way.
   In particular, we went  
     beyond previous works by
computing explicitly the evolution of the {\em second-order}
adiabatic and entropy components on {\em large scales},
in the case in which the scalar fields have non-canonical
kinetic terms. The second-order adiabatic component is, in this
limit, governed by a local first-order (in time) evolution
equation, sourced by terms depending on the second-order entropy
perturbation as well as, quadratically, on the first-order entropy
perturbation. Both first and second order entropy perturbations
satisfy a second order (in time) evolution equation and the full
system of equations, valid  on large scales, is thus closed.  The final
 system of equations enables one to clearly identify
  new effects due to the non-canonical structure of the  scalar fields Lagrangian.
A representative example is equation (\ref{zeta2_evol}),
 that governs the second-order evolution equation for the curvature perturbation on large scales.
This quantity is sourced by combinations of first and second-order 
entropy perturbations, with coefficients depending on the form   
of the kinetic terms.
As discussed in Sec. \ref{last}, in some situations, cancellations can occur, reducing the   
size of some coefficients with respect to the others. In these
cases, the curvature pertubation would be sensitive only to some type
of terms, with important consequences for the evolution of second-order perturbations and thus for non-Gaussianities. 
 All the results derived on large scales in Sec.~\ref{sec:secondorder} are applicable to the interesting scenario of multifield DBI inflation of the type studied in \cite{Langlois:2008wt,Langlois:2008qf}.

\smallskip

In conclusions, we have shown that 
our results, and more in general
 the covariant approach to the dynamics of fluctuations, 
 can be important to analyze the evolution of 
perturbations in multi-field  models of inflation
inspired by high-energy physics or string
theory. This interesting  subject will offer,
in the near future,  unique  opportunities 
to allow comparisons between predictions of 
 high-energy physics motivated inflationary models, and
 observations of the cosmic microwave background radiation.

\vspace{0.9cm}
\centerline{\bf Acknowledgments}
\vspace{0.4cm}

\noindent
We thank David Langlois, Dani\`ele A. Steer and Filippo Vernizzi for very useful discussions and for their careful reading of the draft. We also would like to thank Jean-Luc Lehners for pointing out a typo in Eq.~(\ref{delta_theta_2}).
Most of this work has been done while G.~T. was partially supported by
  MEC and FEDER under grant FPA2006-05485, by CAM under grant HEPHACOS P-ESP-00346, and by the UniverseNet network
   (MRTN-CT-2006-035863). He thanks the kind hospitality
   of APC, Paris,
    where
   this work was initiated.

\appendix
\section{Useful identities in a two-field system}

\subsection{Covariant identities}

\bea
 D_a \dot \sigma &=&  \dot \sigma_a +\ddot\sigma u_a -\dot \theta 
  s_a \,=\,  \dot{\sigma}_a^ \perp  -\dot \theta 
  s_a \,,\\
  D_a X &=&\dot{\sigma}\, D_a \dot \sigma -
\frac12 D_a \Pi \label{daX}
 \,,\\
  D_a P_{,X} &=& P_{,X X}\,D_a X + P_{,X \sigma}\,\sigma_a^\perp
  + P_{,X s}\,s_a  \,,\label{dapX}\\
 \dot{P}_{,X} &=& P_{,X X}\,\left[ 
\dot{\sigma}\, \ddot \sigma -\frac12 
\dot \Pi
 \right] + P_{,X \sigma}\,\dot{\sigma}
   \,,\\
    D_a P_{,XX} &=& P_{,X X X}\,D_a X + P_{,XX \sigma}\,\sigma_a^\perp
  + P_{,X Xs}\,s_a  \,,\label{dapXX}
  \, \\
     D_a P_{,X \sigma} &=& P_{,X X \sigma}\,D_a X + P_{,X\sigma \sigma}\,\sigma_a^\perp
  + P_{,X \sigma s}\,s_a  \,,\label{dapXsi}
  \, \\
    D_a \dot{P}_{,X} &=& \left(
    D_a P_{,XX}
    \right)
    \,
    \left[ 
\dot{\sigma}\, \ddot \sigma -\frac12 
\dot \Pi
 \right] +
D_{a}\,\left(P_{,X \sigma}\,\dot{\sigma} \right)
\nonumber \\
 &+&
 P_{,X X}\,\left[ 
\left(D_a \dot{\sigma}\right)\, \ddot \sigma -\frac12 
D_a\,
\dot \Pi
 \right] + P_{,XX} \dot{\sigma}\,\left[ 
 \ddot{\sigma}_a^\perp-\ddot{\theta} s_a -\dot{\theta} \dot{s}_a
 \right]\label{dotcompl}
\eea

\subsection{Background identities}
\beq
H'=-4 \pi G \bar P_{,X}  \sib^{\prime 2}.
\eeq

\bea
\bar P_{,\sigma}' \tackl = \tackr \sib'  \left( \bar P_{,\sigma\sigma}+\sib'' \bar P_{,X \sigma} \right)
+\thetab' \bar P_{,s},
\\
\bar P_{,s}' \tackl = \tackr \sib'  \left( \bar P_{,\sigma s}+\sib'' \bar P_{,X s} \right)
-\thetab' \bar P_{,\sigma},
\\
\bar P_{,\sigma s}' \tackl =  \tackr \sib'  \left( \bar P_{,\sigma \sigma s}+\sib'' \bar P_{,X \sigma s} \right)+\thetab' \left( \bar P_{,s s}-\bar P_{, \sigma \sigma} \right),
\\
\bar P_{,ss}'\tackl = \tackr \sib'  \left( \bar P_{,\sigma s s}+\sib'' \bar P_{,X s s} \right)-2 \thetab'   \bar P_{,\sigma s},
\\
\bar P_{,X}' \tackl = \tackr \sib'  \left( \bar P_{, X\sigma}+\sib'' \bar P_{,X X} \right),
\\
\bar P_{,XX}' \tackl = \tackr \sib'  \left( \bar P_{, XX\sigma}+\sib'' \bar P_{,XX X} \right),
\\
\bar P_{,X \sigma}' \tackl = \tackr \sib'  \left( \bar P_{,X \sigma\sigma}+\sib'' \bar P_{,X X \sigma} \right)
+\thetab' \bar P_{,Xs},
\\
 \bar P_{,Xs}' \tackl = \tackr \sib'  \left( \bar P_{,X \sigma s}+\sib'' \bar P_{,XX s} \right)
-\thetab' \bar P_{,X\sigma},
\\
.
\eea

\bea
\thetab'\tackl = \tackr\frac{\bar P_{,s}}{ \bar P_{,X} \sib'}, \\
\bar \theta''\tackl = \tackr \frac{1}{\bar P_{,X}} \left(\bar P_{\sigma s}+\sib'' \bar P_{,X s} \right)-\frac{\bar \theta'}{\bar P_{,X} \sib'} \left( \bar P_{,\sigma} +\bar P_{,X \sigma}{\overline \sigma'}^{2}  +\sib'' \frac{\bar P_{,X}}{\bar c_s^2} \right), \label{thetadotbar}
\eea

\subsection{First-order identities}

\bea
\delta e_{\sigma}^I \tackl = \tackr \frac{1}{\sib'} \left(\delta s'+\bar \theta' \delta \sigma \right) \bes^I-\bar \Gamma^{I}_{JK} \besi^J \left(\besi^K \delta \sigma+\bes^K \delta s \right), \\
\delta e_{s}^I \tackl = \tackr -\frac{1}{\sib'} \left(\delta s'+\bar \theta' \delta \sigma \right) \besi^I-\bar \Gamma^{I}_{JK} \bes^J \left(\besi^K \delta \sigma+\bes^K \delta s \right), \\
\delta(\dot\sigma) \tackl = \tackr \sif' - \bar \theta' \sf
-\sib'A \, , \\
\delta P_{,s} \tackl = \tackr \bar P_{,s\sigma} \sif + \bar
P_{,ss} \sf +\bar P_{,X s} \left( \sib'  \delta(\dot\sigma)-\frac 1 2 \delta \Pi \right)-\frac{\bar P_{,\sigma}}{\sib'} (\sf' +\bar \theta'
\sif), \\
\delta P_{,Xs} \tackl = \tackr \bar P_{,Xs\sigma} \sif + \bar
P_{,Xss} \sf +\bar P_{,XX s} \left( \sib'  \delta(\dot\sigma)-\frac 1 2 \delta \Pi \right)-\frac{\bar P_{,X \sigma}}{\sib'} (\sf' +\bar \theta'
\sif), \\
\delta P_{ss} \tackl = \tackr \bar P_{,ss\sigma} \sif + \bar
P_{,sss} \sf  + \bar P_{,Xs s} \left( \sib'  \delta(\dot\sigma)-\frac 1 2 \delta \Pi \right)- 2 \frac{\bar P_{,s\sigma}}{\sib'} \left(\sf' +\bar
\theta' \sif \right), \label{delta_Vss} \\
\delta P_{,X} \tackl = \tackr \bar P_{,X\sigma} \sif + \bar
P_{,Xs} \sf +\bar P_{,XX } \left( \sib'  \delta(\dot\sigma)-\frac 1 2 \delta \Pi \right), \\
\delta  \dot P_{,X} \tackl = \tackr  (\delta P_{,X})'-A \bar P_{,X}  \,.
\eea

\end{document}